\documentclass[peerreview, 11pt]{IEEEtran}
\usepackage{epsfig,amssymb,amsmath,cite,theorem}
\usepackage{setspace}

\theorembodyfont{\normalfont}
\newtheorem{theorem}{Theorem}
\newtheorem{lemma}{Lemma}

\newtheorem{proposition}{Proposition}

\newtheorem{remark}{Remark}

\newtheorem{example}{Example}

\doublespace

\newcommand{\bm}[1]{\mbox {\boldmath $#1$}}

\newcommand{\E}{{\sf E}}
\newcommand{\C}{{\mathcal C}}
\newcommand{\cth}{c_{\star}}
\newcommand{\ubth}{c_{0}}
\newcommand{\snr}{{\sf SNR}}

\begin{document}
\renewcommand{\textfraction}{0}

\title{\LARGE{\rm Incremental Redundancy Cooperative Coding for Wireless
Networks:\\[2mm]
Cooperative Diversity, Coding, and Transmission Energy Gain}}
\author{
Ruoheng Liu, Predrag Spasojevi\'c, and Emina Soljanin
\thanks{The work was supported in part by NSF Grant CCR-0205362 and SPN-0338805.
The material in this paper was presented in part at the 41st
Allerton Conference on Communication, Control, and Computing,
Urbana, IL, USA, Oct., 2003, IEEE 6th Workshop on Signal Processing
Advances in Wireless Communications, NY, USA, June 2005, and IEEE
Information Theory Workshop, Chengdu, China, Oct., 2006.}
\thanks{R.\ Liu and P.\ Spasojevi\'c are with WINLAB, Electrical and Computer Engineering Department,
Rutgers University, North Brunswick, NJ 08902 USA. Email:
\{liurh,spasojev\}@winlab.rutgers.edu}
\thanks{E.\ Soljanin is with Mathematical Sciences Research Center, Bell Labs, Lucent, Murray Hill, NJ 07974 USA.
Email: emina@lucent.com}}
\maketitle \thispagestyle{empty}

\vspace{10mm}

\renewcommand{\baselinestretch}{1.5}

\begin{abstract}
We study an \emph{incremental redundancy} (IR) cooperative coding scheme for
wireless networks. To exploit the distributed spatial diversity benefit we
propose a cluster-based collaborating strategy for a quasi-static Rayleigh
fading channel model. Our scheme allows for enhancing the reliability
performance of a direct communication over a single hop. The collaborative
cluster consists of $M-1$ nodes between the sender and the destination. The
transmitted message is encoded using a mother code which is partitioned into
$M$ blocks each assigned to one of $M$ transmission slots. In the first slot,
the sender broadcasts its information by transmitting the first block, and its
helpers attempt to relay this message. In the remaining slots, each of
left-over $M-1$ blocks is sent either through a helper which has successfully
decoded the message or directly by the sender where a dynamic schedule is based
on the ACK-based feedback from the cluster. By employing powerful \emph{good
codes} including turbo, low-density parity-check, and repeat-accumulate codes,
our approach illustrates the benefit of collaboration through not only a
cooperation diversity gain but also a coding advantage. The basis of our error
rate performance analysis is the threshold behavior of good codes. We derive a
new \emph{simple code threshold} for the Bhattacharyya distance based on the
modified Shulman-Feder bound and the relationship between the Bhattacharyya
parameter and the channel capacity for an arbitrary binary-input
symmetric-output memoryless channel. The study of the diversity and the coding
gain of the proposed scheme is based on this new threshold. An average
frame-error rate (FER) upper bound and its asymptotic (in SNR) version are
derived as a function of the average fading channel SNRs and the code
threshold. Based on the asymptotic bound, we investigate both the diversity,
the coding, and the transmission energy gain in the high and moderate SNR
regimes for three different scenarios: transmitter clustering, receiver
clustering, and cluster hopping. Furthermore, given a geometric distance
profile of the network, we have observed that the energy saving of the IR
cooperative coding scheme is {\it universal} for all good code families.
\end{abstract}


\newpage
\doublespace

\section{Introduction}\label{sec:intro}

To overcome fading, wireless networks employ various diversity techniques,
e.g., channel interleavers, multiple antennas, frequency hopping, etc. In
\cite{send:erkip:aazh}, Sendonaris, Erkip, and Aazhang have proposed the
so-called {\it user-cooperation diversity} where users partner in sharing their
antennas and other resources to create a virtual array through distributed
transmission and signal processing. Only limited time and frequency diversity
is available when the transmitted signal experiences a quasi-static
frequency-flat fading and the receiver has a strict decoding delay constraint
(as in speech and video transmission). In this case, introducing spatial
diversity through cooperation is especially beneficial.

Motivated by the increasing importance of cooperative communication in wireless
networks, a large of valuable work has gone into designing the collaborative
protocols, assessing the information-theoretic limits, and evaluating the
cooperative benefit in the past few years. An information-theoretic analysis of
several cooperative protocols has been reported in \cite{ltw-IT,lw-IT} based on
{\it repetition-coding} and {\it space-time-coding}. In an independent,
relevant work \cite{hunt-IT}, the authors have studied the outage probability
and the asymptotic error probability performance calculations of a two-user
incremental redundancy coding scheme based on Gaussian codebooks. An practical
coding strategy for two-user collaborating transmission based on rate
compatible punctured convolutional codes has been considered in
\cite{stef:erkip,hunt-TWCOM}. Cooperation schemes employing distributed turbo
codes have been studied in \cite{zhao:vale,rpe-coop,jana:heda:hunt:nosra},
where \cite{rpe-coop} has focused on the asymptotic error performance analysis.

In this paper, we study an \emph{incremental redundancy} (IR) cooperative
coding scheme for a multiple-helper wireless network. To exploit the spatial
diversity benefit we propose a cluster-based collaborating strategy for a
quasi-static Rayleigh fading channel model and based on a network geometric
distance profile \cite{rpe-spawc2005}. The cluster-based collaborating strategy
enhances the reliability performance of a direct communication over a single
hop. A collaborative cluster consists of $M-1$ nodes assisting the sender in a
two-hop strategy. The transmitted message is encoded using a mother code which
is partitioned into $M$ blocks each assigned to one of $M$ transmission slots.
In the first slot, the sender broadcasts own information by transmitting the
first block, and its helpers attempt to decode the message. In the remaining
slots, each of the left-over $M-1$ blocks is sent either through a helper which
has successfully decoded the message or directly by the sender based on an
acknowledgement (ACK) driven a dynamic schedule. By employing the powerful
\emph{good codes} \cite{mack,rpe-good-IT}, e.g., turbo, low-density
parity-check (LDPC), and repeat-accumulate (RA) codes, whose performance is
characterized by a threshold behavior, our approach illustrates the benefit of
collaboration through not only a cooperation diversity but also a coding
advantage.

To investigate benefits of the proposed IR cooperative coding scheme, we
evaluate its frame error rate (FER) performance for a quasi-static
frequency-flat Rayleigh fading channel. The basis of our error rate performance
analysis is the so-called \emph{code threshold} of a good code ensemble. Our
focus is on the code threshold for which the {\it maximum likelihood} (ML)
decoding word error probability vanishes whenever the channel quality is above
the threshold. In fact, we propose a new Bhattacharyya distance \emph{simple
code threshold} for parallel channel transmissions based on the modified
Shulman-Feder reliable channel region \cite{shul:fede,mill:burs,rpe-IT-II} and
the relationship between the Bhattacharyya parameter and the channel capacity
of an arbitrary binary-input symmetric-output memoryless (BISOM) channel
\cite{rp-ITW2006}. The derived threshold characterizes a reliable communication
condition for parallel channel transmissions with a single constraint in terms
of the average Bhattacharyya parameter. Compared with the union Bhattacharyya
(UB) code threshold introduced in \cite{huj:mceli}, the new code threshold is
almost tighter $1~dB$ for an example turbo code transmitted over a binary-input
AWGN (BI-AWGN) channel, as illustrated in Example~\ref{ex:tc-cth}.

Next, based on the simple code threshold and the outage concept, we derive a
FER upper-bound which predicts well the fading channel simulation results for
the proposed cooperative coding scheme. Compared with the result in
\cite{hunt-TWCOM} (and the block-fading channel FER bound in \cite{maik:leib}),
the derived FER bound is a function of the single code threshold parameter
instead of a family of weight enumerators. Hence, this simple FER bound is not
only easy to compute but is also insightful in that it allows for a large SNR
asymptotic analysis of the cooperation scheme. In this paper, closed-form
asymptotic FER upper-bounds are obtained for three different scenarios:
transmitter clustering, receiver clustering, and cluster hopping. These bounds
allow for illustrating cooperative diversity and coding gains in the high SNR
regime. Finally, we express FER bounds in terms of the distance profile and
investigate cooperation benefits in terms of the energy efficiency gain for
different collaborative cluster sizes and normalized cluster distances.

The paper is organized as follows. We describe the system model in
Sec.~\ref{sec:sys}. We study the threshold behavior of good code ensembles over
BISOM channels in Sec.~\ref{sec:pre}. An IR cooperative coding scheme is
introduced in Sec.~\ref{sec:cptc}. We derive an upper bound on the scheme FER
in Sec.~\ref{sec:fer}, and its asymptotic versions for different cooperation
scenarios in Sec.~\ref{sec:limit}. Simulation results and the collaborative
cluster design are discussed in Sec.~\ref{sec:sd}. Finally, we summarize our
results in Sec.~\ref{sec:con}.

\section{System Model}\label{sec:sys}
In this section, we introduce a collaborative cluster and describe a fading
channel model based on the geometric distance profile of the network.

\subsection{Collaborative Cluster}

\begin{figure}[hbt]
  \centerline{\includegraphics[width=0.6\linewidth,draft=false]{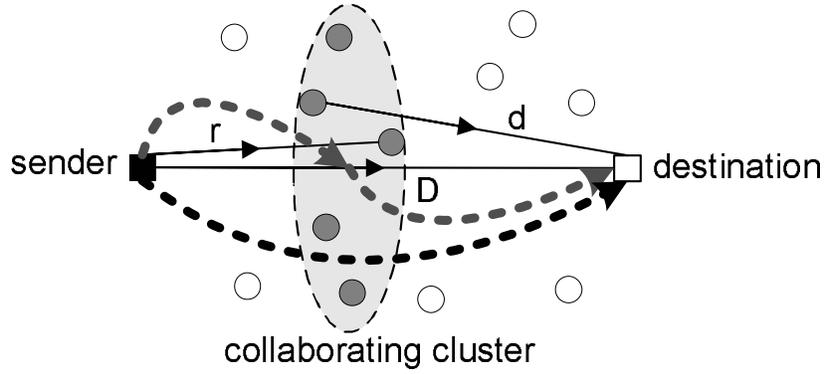}}
  \caption{\small Cooperative routing in wireless network with a geometric distance profile} \label{fig:dis}
\end{figure}

As shown in Fig.~\ref{fig:dis}, we consider a single sender-destination pair
and several intermediate nodes. A group of $M-1$ nodes forms a {\it
collaborative cluster} $\mathcal S$ which assists the sender in a two-hop
scheme. The sender serves as the regional broadcast node, and each cluster
member, termed \emph{helper}, attempts to relay the package to the destination.
For notational convenience, let Node $0$ denote the sender, Node
$m$, for $m=1,\dots,M-1$, denote a helper, and Node $M$ denote the
destination.

\begin{remark}
The proposed cluster-based collaborating strategy can be embedded into an
existing noncooperative route. For example, the routing metric most commonly
used in existing ad hoc routing protocols is the minimum hop-count, which
results in a long distance traveled by each hop \cite{zhao05:routing}. Hence,
there may exist several intermediate nodes between the (hop) sender and the
(hop) destination for each non-cooperation hop. This is one of the motivations
for the IR cooperative coding scheme proposed in this paper, namely, the
cooperative scheme is a supplement to current routing algorithms for overcoming
the ``dynamic'' deep fade.
\end{remark}

\subsection{Channel Model}
We consider the quasi-static frequency-flat Rayleigh fading channel
model \cite{bigl:proa:sham} where the fading coefficient is random
but invariant during the transmission interval $T$. The
discrete-time channel model is
\begin{align}
y_{i,j}  &= d_{i,j}^{-L/2}\, a_{i,j}\,
x_{i}+z_{i,j}~~\text{for}~i\in\{0,\dots,M-1\}~\text{and}~
j\in\{1,\dots,M\} \label{eq:chmo}
\end{align}
where $x_{i}$ is the signal transmitted by Node~$i$, $L$ is the path loss
exponent, $d_{i,j}$, $a_{i,j}$, and $z_{i,j}$ are the distance, fading
coefficient, and background noise between Nodes~$i$, $j$, respectively, and
$y_{i,j}$ is the signal at Node~$j$ received from Node~$i$. We will focus on
the transmitted signal alphabet is binary, i.e.,
$x_{i}\in\bigl\{+\sqrt{E},-\sqrt{E}\bigr\}$, where $E$ is the transmitted
symbol energy which is identical for all cluster nodes and the sender, and the
symbol ``$+\sqrt{E}$'' (and ``$-\sqrt{E}$'') represents the coded symbol
``$0$'' (and ``$1$''). We assume that $z_{i,j}$ is modeled as the mutually
independent additive Gaussian noise ${\mathcal N}(0,1/2)$, and
$\nu_{i,j}=|a_{i,j}|^2$ is the exponentially distributed ``channel power'' with
mean $1$. Then, the average and instantaneous SNRs of the signal at Node $j$
received from Node $i$ can respectively be expressed as
$$\snr_{i,j}\triangleq E\cdot d_{i,j}^{-L}, ~~\text{and}~~
\theta_{i,j}\triangleq \nu_{i,j}\cdot \snr_{i,j}.$$ We further
assume that decoding is done with the knowledge of the fading
coefficients. Note that, for a given fading coefficient $a_{i,j}$,
the channel (\ref{eq:chmo}) is a BI-AWGN channel. This motivates us
to study the error performance of binary good code ensembles
transmitted over BISOM channels in the next section.

\section{Threshold Based Performance Analysis for BISOM channels}\label{sec:pre}

In this section, we introduce the notation and the preliminary material on
BISOM channel measures and the weight spectrum of code ensembles. Next, we
study the threshold behavior of good code ensembles and derive a new simple
code threshold for parallel channel transmissions. Results given in this
section are used as the foundation for the error performance analysis of the IR
cooperative coding scheme.

\subsection{Channel Capacity and Bhattacharyya Parameter of BISOM Channels}

Here, we consider a binary input memoryless channel with the output
alphabet $\mathcal Y$ and transition probabilities $p(y|0)$ and
$p(y|1)$, $y\in\mathcal Y$. We say that the channel is
\emph{symmetric} if $p(y|0)=p(-y|1)$. We first study the channel
capacity and the Bhattacharyya parameter for BISOM channels. The
Bhattacharyya parameter is widely used to characterize the
``noisiness'' of the channel (e.g., see \cite{huj:mceli,rpe-IT-II}).
The capacity of a BISOM channel is achieved by the uniform input
distribution and can be expressed in terms of $p(y|0)$ as follows
\begin{align}
C&= \frac{1}{2}\Biggl[\sum_{y\in{\mathcal Y}}p(y|0)
\log_2\frac{p(y|0)}{p(y)}+ \sum_{y\in{\mathcal Y}}p(y|1)
\log_2\frac{p(y|1)}{p(y)}\Biggr]\nonumber\\
&= \E\left[\log_2\frac{p(Y|0)}{p(Y)}\Biggl|\,0\right] \nonumber\\
&=1-\E\left[\log_2\frac{p(Y|0)+p(-Y|0)}{p(Y|0)}\Biggl|\,0\right].
\label{eq:cd}
\end{align}
Similarly, the Bhattacharyya parameter $\gamma$ is
\begin{align}
\gamma &=\sum_{y\in{\mathcal
Y}}\sqrt{p(y|0)p(y|1)}=\E\left[\sqrt{\frac{p(-Y|0)}{p(Y|0)}}\Biggl|\,0\right].
\label{eq:bndw}
\end{align}
We also consider the {\it cutoff rate} for the BISOM channel,
\begin{align}
R_0\triangleq 1-\log_2(1+\gamma).\label{eq:r0d}
\end{align}
Now, we establish a general relationship among these three
information-theoretic quantities in the following lemma.
\begin{lemma}{\label{lem:BI}}
Let $B\triangleq 1-\gamma$ denote the {\it Bhattacharyya rate}. For
a BISOM channel with a transition probability $p(y|0)$, the channel
capacity, the Bhattacharyya rate, and the cutoff rate satisfy
\begin{align}
C\ge B \ge R_0.
\end{align}
\begin{proof}
The proof in Appendix~\ref{app:BI} follows the Jensen's inequality.
\end{proof}
\end{lemma}
In Lemma~\ref{lem:BI}, we propose a new channel quality measure called
Bhattacharyya rate which is between the channel capacity and the cutoff rate.
\begin{figure}[hbt]
\noindent
  \begin{minipage}[b]{.5\linewidth}
 \centerline{\includegraphics[width=0.9\linewidth,draft=false]{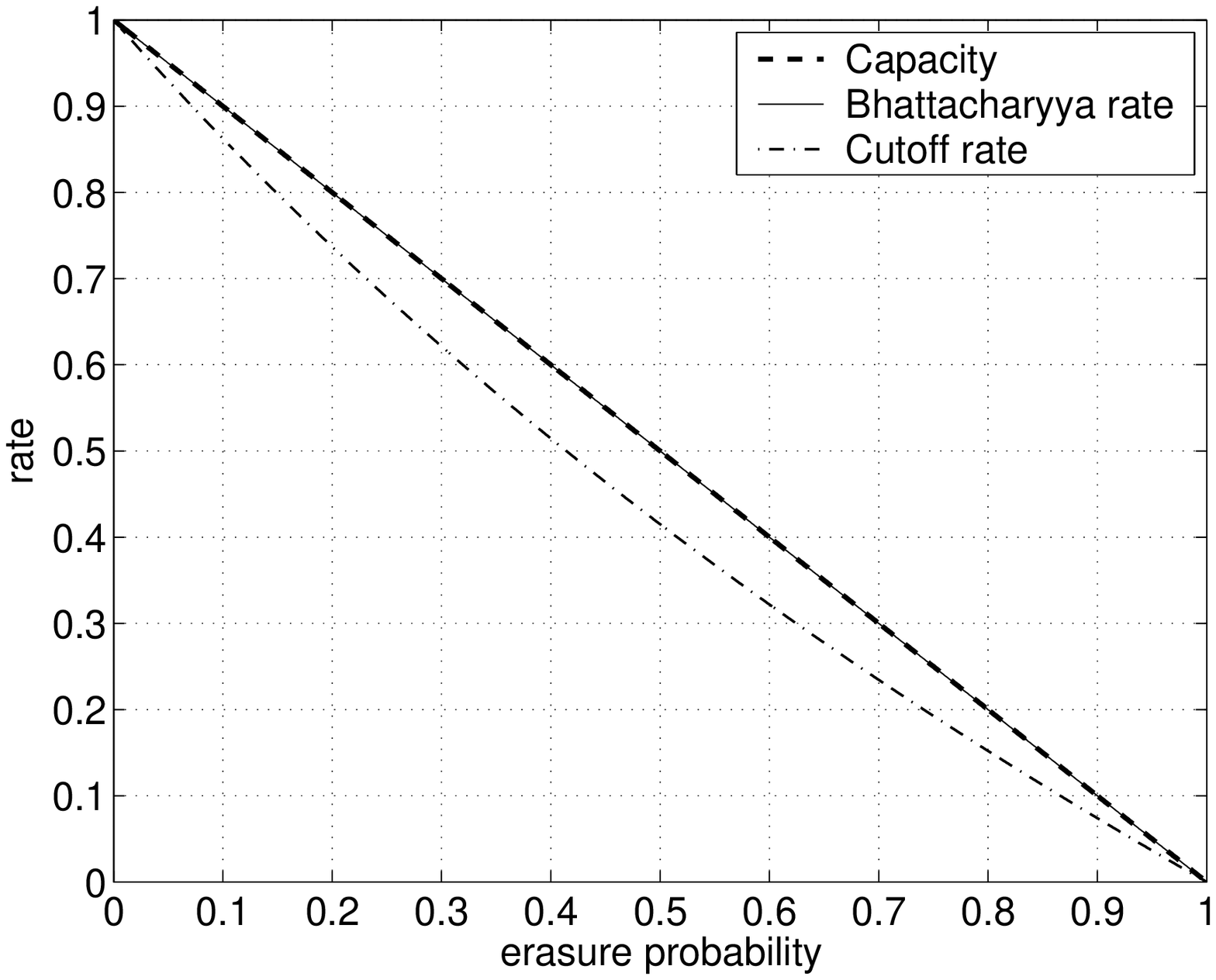}}
    \centerline{\mbox{\small a. BEC }}
  \end{minipage}
\hfill
  \begin{minipage}[b]{0.5\linewidth}
  \centerline{\includegraphics[width=0.9\linewidth,draft=false]{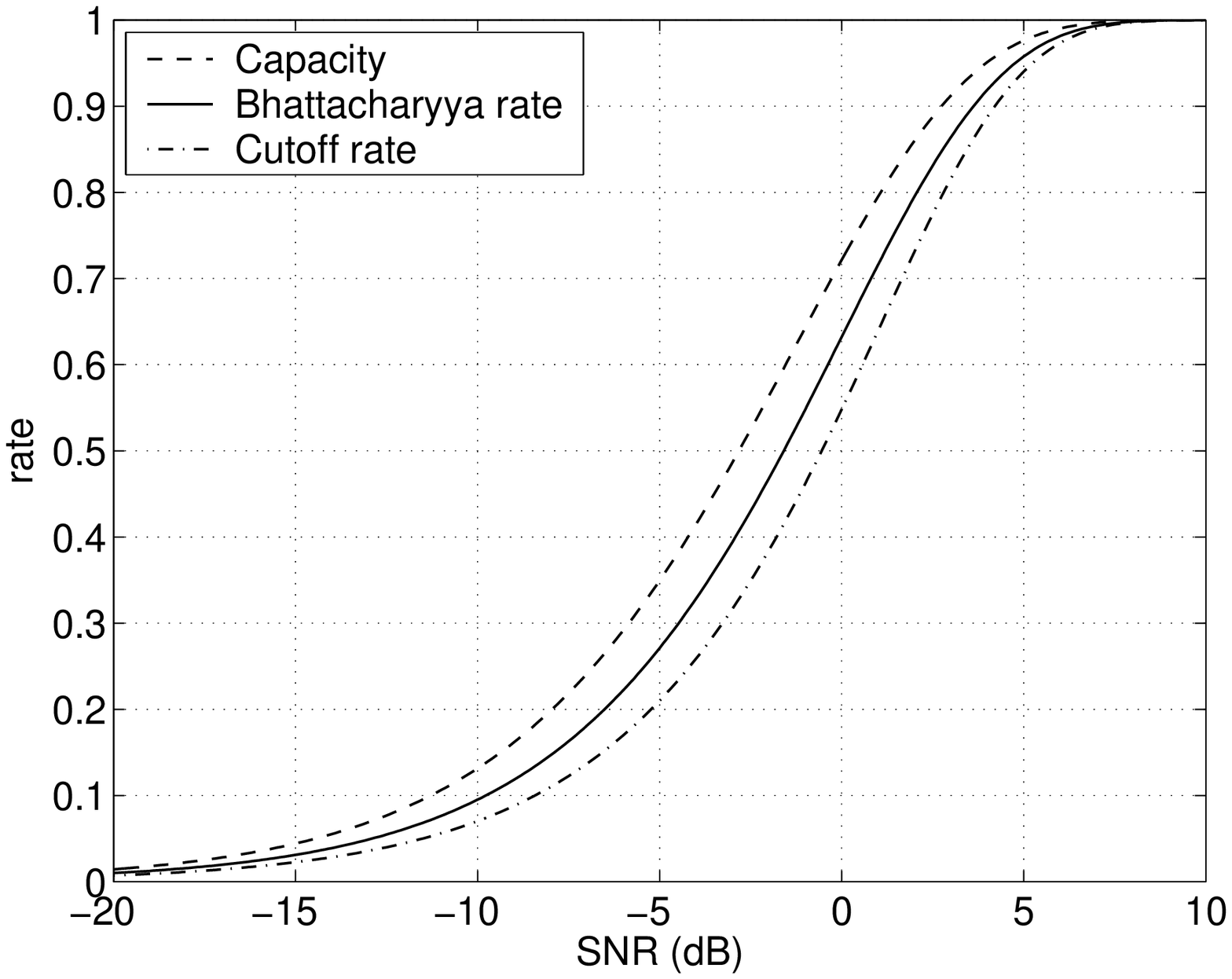}}
    \centerline{\mbox{\small b. BI-AWGN channel}}
  \end{minipage}
 \caption{\small Capacity, Bhattacharyya rate, and cutoff rate for
  BECs and BI-AWGN channels} \label{fig:cbc}
\end{figure}
In particular, the Bhattacharyya rate is equal to the channel capacity for the
case of the {\it binary erasure channel} (BEC). For the case of a BI-AWGN
channel with the SNR $\lambda$,
\begin{align}
C(\lambda)&=1-
\frac{1}{\sqrt{\pi}}\int_{-\infty}^{+\infty}e^{-(y-\sqrt{\lambda})^2}
\log_2\bigl(1+e^{-4y\sqrt{\lambda}}\bigr)\,dy \label{eq:C-awgn}\\
B(\lambda)&=1-e^{-\lambda}\\
R_0(\lambda)&=1-\log_2 (1+e^{-\lambda}). \label{eq:awgn}
\end{align}
Clearly, for a BI-AWGN channel, the numerical calculation of the Bhattacharyya
rate is much easier than the computation of the channel capacity which requires
integration form from $-\infty$ to $\infty$. The channel capacity, the
Bhattacharyya rate, and the cutoff rate are illustrated in Fig.~\ref{fig:cbc}
for BECs and BI-AWGN channels.

\subsection{Weight Spectrum Properties of Good Code Ensembles}

In this paper, we consider good binary linear code ensembles
\cite{mack,rpe-good-IT} whose performance is characterized by a threshold
behavior. The class of good codes includes turbo, LDPC, and RA codes, and some
recent variations. Due to the outstanding performance of such random-like
codes, we employ these in our IR cooperative coding scheme. In order to
evaluate the error performance of the cooperation scheme, we study code
thresholds based on the weight spectrum of code ensembles. The weight spectrum
for various good code ensembles were studied in, e.g.,
\cite{gallager_2,huj:mceli,divs,sason}. In this subsection, we briefly
introduce some notation on weight spectra of binary linear code ensembles,
review the UB code threshold proposed in \cite{huj:mceli}, and describe a
relationship between the UB code threshold and the code rate.

For a given binary linear code ensemble $[\C]$ of length $N$ and rate $R$, let
$A^{[\C]}_{h}$ denote the average number of codewords of Hamming weight $h$,
termed the {\em average weight enumerator} (AWE). Let $\delta\triangleq h/N$
denote the normalized Hamming weights, then the {\it asymptotic normalized
exponent of the weight spectrum} with respect to the codeword length is defined
as
\begin{align}
r^{[\C]}(\delta)\triangleq \limsup_{N\rightarrow\infty} \frac{\ln
A^{[\C]}_h}{N}
\end{align}
where the superscript $[\C]$ denotes a binary code ensemble\footnote{We shall
use the symbol $\C$ to denote a binary codeword, $\{\C\}$ to denote a codebook,
and $[\C]$ to denote a binary code ensemble.}. To simplify notation, hereafter,
we drop the superscript $[\C]$ from code parameters, e.g., in
$r^{[\C]}(\delta)$ and $A^{[\C]}_{h}$, when we do \emph{not} specify a
particular code ensemble. Now, the {\it Shulman-Feder} (SF) distance
\cite{shul:fede}
\begin{align}
\xi \triangleq \sup_{0<\delta \le 1} \bigl[r
(\delta)-r^{[\mathcal{RB}]}(\delta)\bigr]\log_2 e
\end{align}
measures the weight spectrum distance between the random binary code
ensemble $[\mathcal{RB}]$ and the code ensemble $[\C]$, where
$$r^{[\mathcal {RB}]}(\delta)=H(\delta)-(1-R)\ln2$$
is the asymptotic normalized exponent of the weight spectrum for the random
binary code ensemble and
$$H(\delta)\triangleq-\delta\ln(\delta)-(1-\delta)\ln(1-\delta)$$
is the {\it binary entropy function}. Following \cite{huj:mceli}, we define the
UB code threshold of a code ensemble $[\C]$ as
\begin{align}
\ubth\triangleq \sup_{0 <\delta \le1}\frac{ r(\delta)}{\delta}.
\label{eq:c0}
\end{align}

In this paper, we consider a family of good code ensembles whose
weight spectra satisfies the following condition:
\begin{enumerate}
  \item For a given binary code ensemble $[\C]$, there exists a sequence of integers
$D_N$($<N$) such that $D_N\rightarrow\infty$ and
\begin{equation}
\lim_{N\rightarrow\infty}\sum_{h=1}^{D_N} A_{h}=0; \label{eq:lwawe}
\end{equation}
  \item and the UB code threshold $\ubth$ is finite.
\end{enumerate}

Finally, we establish a relationship between the code rate $R$ and the
threshold $\ubth$ in the following lemma.
\begin{lemma}{\label{lem:rate}}
For a good binary linear code ensemble $[\C]$ of rate $R$, the UB code
threshold $\ubth$ is lower bounded by the following function of $R$,
\begin{align}
\ubth \ge -\ln (1-R). \label{eq:lm-rate}
\end{align}
\begin{proof}
We show the proof in Appendix~\ref{app:rate}.
\end{proof}
\end{lemma}

\subsection{Threshold Behavior of Good Codes for BISOM Channels} \label{subsec:tctb}

The basis of our analysis is the threshold behavior of good codes. Here, we
first review the work in \cite{huj:mceli,erp}, which has studied the error
performance of good code ensembles (e.g., turbo codes) based on the UB code
threshold $\ubth$. Next, we introduce a tighter Bhattacharyya distance code
threshold $\cth$ (compared with $\ubth$) under ML decoding. Based on the new
threshold $\cth$, we derive a coding theorem for good code ensembles whose
transmission takes place over independent parallel channels.

\subsubsection{UB Code Threshold}

In \cite{huj:mceli}, Jin and McEliece have shown that, for a
binary-input memoryless channel, if
\begin{align}
\gamma<\exp(-\ubth), \label{eq:condc0}
\end{align}
the average ML decoding word error probability approaches zero. Inequality
(\ref{eq:condc0}) describes a reliable communication condition for the code
ensemble $[\C]$ based on a single channel parameter $\gamma$ and a single code
parameter $\ubth$, where the Bhattacharyya parameter $\gamma$ represents the
``noisiness'' level of the channel and the UB code threshold $\ubth$
characterizes weight spectrum properties of the code ensemble. This result is
based on the classical union bound. Hence, the threshold $\ubth$ is loose.
Based on $\ubth$ and the random assignment method (described below), we have
studied this bound for parallel BISOM channels \cite{erp,rpe-IT-II}.

\subsubsection{Simple Code Threshold}

Following the approach in \cite{mill:burs}, for a given good code
ensemble $[\C]$ of rate $R$, we partition the normalized Hamming
weights $\delta$ into two disjoint subsets,
\begin{align*}
\Psi(P)&\triangleq\{\delta: 0<\delta\le 0.5-P \text{ or } 0.5+P <
\delta \le
1\} \nonumber\\
\Psi^{c}(P)&\triangleq\{\delta: 0.5-P <\delta \le 0.5+P\}, \text{
for } P\in[0,\,0.5).
\end{align*}
Next, we define a new Bhattacharyya distance code threshold by
optimizing the weight partition parameter $P$ as follows
\begin{align}
\cth \triangleq  \min_{0\le P< 0.5}
    \left\{c_{P} :~c_{P}  \ge - \ln(1-R- \xi_{P} )\right\} \label{eq:ubth}
\end{align}
where
\begin{align}
c_P\triangleq \sup_{\delta \in \Psi(P)}\frac{r(\delta )}{\delta}
\label{eq:cp}
\end{align}
denotes the restriction UB code threshold corresponding to the
weight subsets $\Psi(P)$, and
\begin{align}
\xi_{P}\triangleq \left\{
\begin{array}{ll}
   \displaystyle{\sup_{\delta \in \Psi^{c}(P)}\bigl[r(\delta)-r^{[\mathcal{RB}]}(\delta)\bigr]\log_2
e}, & P>0 \\
      0, & P=0.
\end{array}\right. \label{eq:xip}
\end{align}
denotes the restriction SF distance corresponding to the weight
subsets $\Psi^{c}(P)$. Note that, in the case $P=0$,
\begin{align*}
c_{P}=\ubth \quad \text{and} \quad \xi_{P}=0.
\end{align*}
Hence, Lemma~\ref{lem:rate} and (\ref{eq:ubth}) imply that the new
code threshold $\cth \le \ubth$.

Now, we consider the average error rate performance of a code ensemble $[\C]$
transmitted over $Q$ parallel channels. Following the random assignment
approach \cite{erp}, we assume that the bits of the transmitted codeword are
randomly assigned to parallel channels so that each bit is transmitted over
Channel $j$ with the a-prior probability $\tau_j$, where
$\sum_{j=0}^{Q-1}\tau_j=1$ and $\tau_j>0$ for $j=0,\dots,Q-1$. We have the
following parallel channel coding theorem for the code ensemble $[\C]$ based on
$\cth$.

\begin{theorem}\label{th:cth}
Let us consider a linear binary code ensemble $[\C]$ whose transmission takes
place over $Q$ independent parallel BISOM channels. Assume that the bits of the
transmitted codeword are randomly assigned to the channels with assignment
rates $\{\tau_0,\dots,\tau_{Q-1}\}$. Let
\begin{align}
\bar{\gamma}\triangleq \sum_{j=0}^{Q-1}\tau_j \gamma_j
\end{align}
be the average Bhattacharyya parameter over $Q$ parallel channels, where
$\gamma_j$ is Bhattacharyya parameter of Channel $j$, for $j=0,\dots,Q-1$. If
\begin{align}
\bar{\gamma} <\exp(-\cth), \label{eq:sth-cd}
\end{align}
then the average ML decoding word error probability
$P_{W}(\bar{\gamma},N)\stackrel{N}{\longrightarrow}0.$

\begin{proof}
The proof is based on the relationship between the channel capacity
and the Bhattacharyya rate in Lemma~\ref{lem:BI}, the lower bound on
the UB code threshold in Lemma~\ref{lem:rate}, and the {\it modified
Shulman-Feder} reliable channel region in Appendix~\ref{app:msf}
(please also refer to \cite{rpe-good-IT} for details). We provide
the proof of Theorem~\ref{th:cth} in Appendix~\ref{app:cth}.
\end{proof}
\end{theorem}
In Theorem~\ref{th:cth}, the reliable communication condition (\ref{eq:sth-cd})
is a simple constraint in terms of the average Bhattacharyya parameter. The new
code threshold allows for characterizing more complex coding schemes (e.g.,
communication over block-fading channel where the error performance requires
averaging over all possible channel realizations) in a simple and more accurate
manner. Hence, we refer to $\cth$ as the \emph{simple} code threshold for
$[\C]$. The above simple code threshold theorem describes the asymptotic result
where we let the codeword length $N$ tend to infinity. The following example
illustrates that the simple code threshold aids in estimating the error
performance of long codes with fixed codeword length\footnote{For typical
practical systems, the codeword length is fairly large \cite{ozar:sham:wyne},
e.g., in CDMA2000 standard \cite{cdma2000}, the encoder allows for a variable
input length up to $K \simeq 20730$.}.
\begin{example}[simple threshold for turbo codes]\label{ex:tc-cth}
Here, we study UB and simple code thresholds of a $R=1/7$ turbo code. The turbo
encoder consists of $J=3$ recursive convolutional encoders with two random
interleavers.
\begin{figure}[hbt]
  \centerline{\includegraphics[width=0.6\linewidth,draft=false]{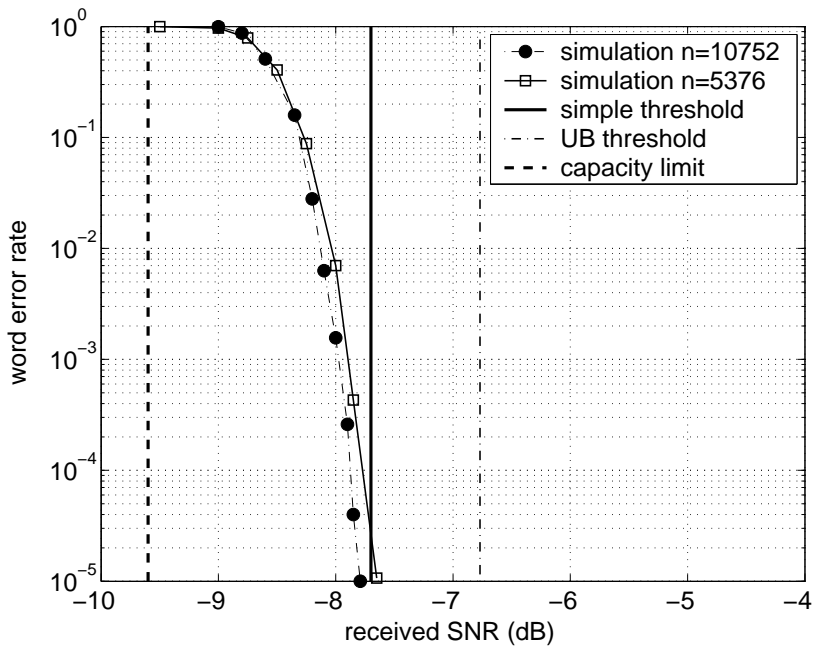}}
  \caption{\small UB and simple code thresholds for an AWGN channel
(turbo code of rate $R=1/7$ and length $N=5376$, $10752$)}
\label{fig:th}
\end{figure}
The component code transfer functions are
$$G_1=(1,\, 13/15,\, 17/15) \text{ and } G_2=G_3=(13/15,\, 17/15).$$
We compute the AWE based on the technique in \cite{divs}. By applying
(\ref{eq:c0}) and (\ref{eq:ubth}), we calculate the UB threshold
$c_0^{[\mathcal{TC}]}\approx 0.21$ and the simple threshold
$c_{\star}^{[\mathcal{TC}]}\approx 0.17$. As shown in Fig.~\ref{fig:th}, we
compare UB and simple thresholds with simulation results for codeword length
$N=5376$ and $N=10752$ under iterative decoding when the turbo codes are
transmitted over a BI-AWGN channel. Fig.~\ref{fig:th} illustrates that the
cliff of WER curves becomes sharp as the codeword length $N$ increases. We
observe that the simple threshold predicts well the cliff of the simulated word
error probability, and that the gap between $c_0^{[\mathcal{TC}]}$ and
$c_{\star}^{[\mathcal{TC}]}$ is almost $1$ dB. We also consider the capacity
limit of the BI-AWGN channel. To this end, we set $C(\lambda)$ in
(\ref{eq:C-awgn}) equal to the code rate $R$, and determine the capacity
threshold SNR. Fig.~\ref{fig:th} illustrates that the capacity threshold may
not predict well the error performance of turbo codes.
\end{example}

\subsubsection{Punctured Code Threshold} \label{subb:pct}

For punctured codes, we may assume that punctured bits are sent to a dummy
memoryless component channel whose output is independent of the input, i.e.,
$\gamma_p=1$, whereas, non-punctured bits are transmitted over the real channel
with the Bhattacharyya parameter $\gamma$. Let the puncturing rate be $1-\tau$,
the average Bhattacharyya parameter is
$\bar{\gamma}=\gamma\cdot\tau+1\cdot(1-\tau)$. Hence, we can rewrite the
reliable communication condition (\ref{eq:sth-cd}) as
\begin{align}
\gamma<\frac{\exp(-\cth)-(1-\tau)}{\tau}. \label{eq:ptcc}
\end{align}
Analogous to Theorem~\ref{th:cth}, we have the following result for
a (randomly) punctured code ensemble.

\begin{theorem} \label{th:pcth}
Consider a good codes ensemble $[\C]$ with a finite code threshold $\cth$
defined in (\ref{eq:ubth}). Assume that the coded symbols are randomly and
independently punctured, so that each bit is punctured with a-priori
probability (punctured rate) $1-\tau$. If $\tau>1-\exp(-\cth)$, there exists a
punctured code threshold
\begin{equation}
\chi(\tau) =  \ln \frac{\tau}{\exp(-\cth)-(1-\tau)}
\label{eq:punctured}
\end{equation}
such that, if
\begin{align}
-\ln \gamma  >\chi(\tau), \label{eq:ccp}
\end{align}
the average ML decoding word error probability approaches zero as
$N\rightarrow\infty$.

\begin{proof}
The proof follows in a straightforward manner from
Theorem~\ref{th:cth} and (\ref{eq:ptcc}).
\end{proof}
\end{theorem}
We note that the left hand side of (\ref{eq:ccp}) is the channel {\it
Bhattacharyya distance}, which is equal to the received SNR for a BI-AWGN
channel.
\begin{figure}[bht]
  \centerline{\includegraphics[width=0.6\linewidth,draft=false]{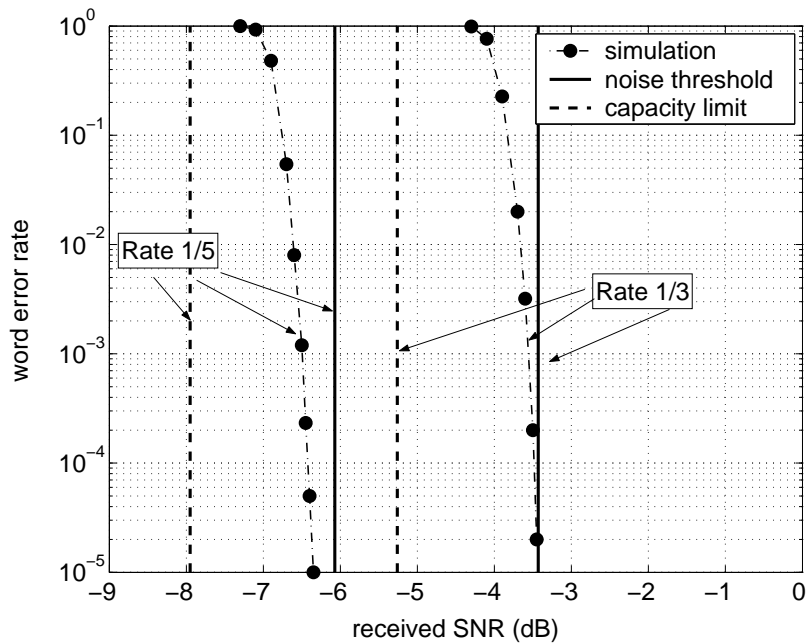}}
  \caption{\small UB and simple code thresholds for an AWGN channel (mother turbo code of $R=1/7$ and $N=26880$)} \label{fig:pth}
\end{figure}
Thus, the simple code threshold is an SNR threshold for an AWGN channel, i.e.,
if the received SNR is larger than the punctured code threshold $\chi(\tau)$,
the average ML decoding word error probability decays to zero as the codeword
length $N$ tends to infinity. We show the SNR threshold behavior of punctured
turbo codes in the following example.
\begin{example}[threshold for punctured turbo codes]\label{ex:pc-cth}
We consider an example punctured turbo code transmitted over a BI-AWGN channel.
The mother turbo code of rate $R=1/7$ and length $N=26880$ is described in
Example~\ref{ex:tc-cth}. We study the word error probability performance of its
punctured versions of rate $R_p=1/5$ (by setting $\tau=5/7$) and $R_p=1/3$ (by
setting $\tau= 3/7$). Based on the simple threshold
$c_{\star}^{[\mathcal{TC}]}\approx 0.17$, we calculate the punctured code
thresholds $\chi(5/7)\approx 0.25$ and $\chi(3/7)\approx 0.45$. As shown in
Fig.~\ref{fig:pth}, we compare the simulation results with punctured thresholds
and capacity SNR limits corresponding to $C=1/5$ and $C=1/3$.
Fig.~\ref{fig:pth} illustrates that our punctured code thresholds are in a very
good agreement with the numerical value observed in the simulation.
\end{example}

\begin{remark} {\it Self-Decodable Condition:} Theorem~\ref{th:pcth}
illustrates that if
\begin{align}
\tau>1-\exp(-\cth), \label{eq:sdcon}
\end{align}
then the punctured code threshold $\chi(\tau)$ exists for a good mother code
ensemble $[\C]$. It implies that (\ref{eq:sdcon}) is a necessary condition
which guarantees that a randomly chosen code (sequence) from the punctured code
ensemble is self-decodable with probability one. Hence, we refer to
(\ref{eq:sdcon}) as a self-decodable condition.
\end{remark}

\section{IR Cooperative Coding Scheme}\label{sec:cptc}

In this section, we introduce an IR coding collaboration scheme. We assume that
the sender and helpers acquire a common time reference and operate in a single
radio frequency band $W$. The medium access control (MAC) scheme is based on a
time-division scheme. The transmission interval $T$ is partitioned into $M$
non-overlapping slots of duration $\tau_0 T,\dots,\tau_{M-1} T$, where
$\tau_j$, for $j=0,\dots,M-1$, is referred to as the {\it assignment rate} for
slot $j$, $\sum_{j=0}^{M-1} \tau_j = 1,$ and the time allocation
$\{\tau_0,\dots,\tau_{M-1}\}$ is predetermined for all nodes.

Each node in the wireless network has an encoder (associated a codebook
$\{\C\}$ of length $N$ and rate $R$), a decoder, and a mapping device. The
sender (Node $0$) encodes the information and obtains a mother codeword $\C$.
The mapping device partitions the codeword $\C$ into $M$ blocks corresponding
to $M$ non-overlapping slots. Block $j$ of length $\tau_j N$ is transmitted in
slot $j$, $j=0,\dots,M-1$. For analysis simplicity, the partition is based on
{\it random assignment} \cite{erp} which can be described using a probabilistic
mapping device which randomly assigns bits of the transmitted codeword to $M$
blocks. More precisely, the random mapper takes a bit of the codeword $\C$ and
sends it to slot $j$ with a predetermined assignment rate $\tau_j$. Bit
assignments are independent and, thus, the expected (and asymptotic as $N
\rightarrow \infty$) number of bits assigned to block $j$ is $\tau_j N$.

After random mapping, each block of bits forms a codeword obtained by
puncturing. Let $\mathcal{C}_{j}$ denote the punctured codeword corresponding
to slot $j$. For a given good (mother) code ensemble $[\C]$ and the (random)
assignment rate $\tau_j$, Theorem~\ref{th:pcth} shows that the punctured code
ensemble exhibits the SNR threshold behavior for an AWGN channel, i.e., if the
received SNR is larger than the punctured code threshold $\chi(\tau_j)$, the
average ML decoding word error probability decays to zero as the codeword
length $N$ tends to infinity. As illustrated in (\ref{eq:punctured}), the
punctured code threshold $\chi(\tau_j)$ can easily be calculated based on the
mother code threshold $\cth$. Hence, this threshold behavior allows for
adaptively scheduling the cooperation in two stages described below.
\begin{figure}[bht]
\vspace{-0.2cm} \noindent
  \begin{minipage}[b]{.45\linewidth}
 \centerline{\includegraphics[width=\linewidth,draft=false]{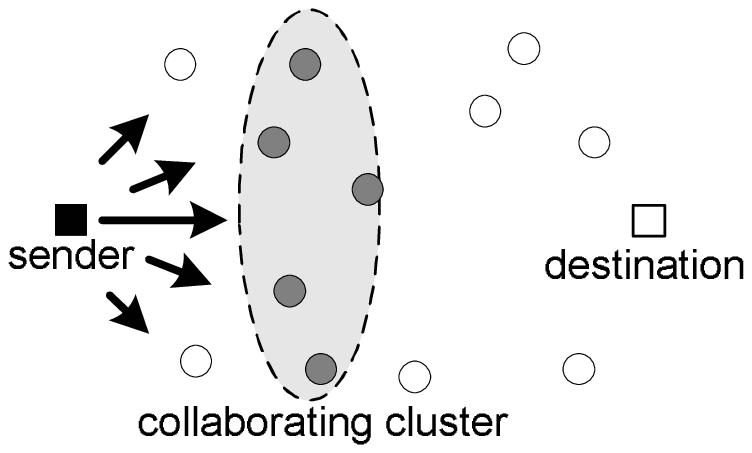}}
\vspace{-0.1cm}
    \centerline{\mbox{\small a. slot $0$: sender broadcasts message }}
  \end{minipage}\hfill
  \begin{minipage}[b]{0.45\linewidth}
  \centerline{\includegraphics[width=\linewidth,draft=false]{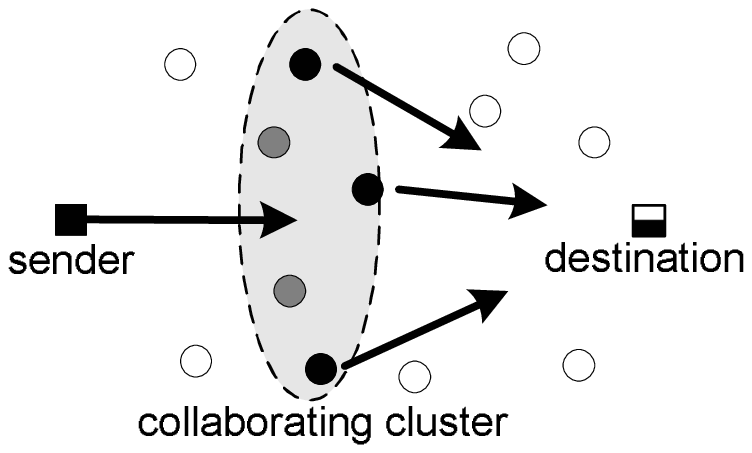}}
    \vspace{-0.1cm}
    \centerline{\mbox{\small b. slot~$1~$-~slot~$M-1$: reliable nodes relay information}}
  \end{minipage}
  \vspace{-0.2cm}
 \caption {\small Cooperative coding scheme}
  \label{fig:sys-model}
\end{figure}

\subsection{Broadcast Stage} \label{subsec:bs}
As shown in Fig.~\ref{fig:sys-model}.a, the sender broadcasts its information
by transmitting $\C_{0}$ during slot~$0$. Each helper listens and attempts to
decode this message. More precisely, helpers estimate the instantaneous
received SNR from the incoming signal and compare the SNR with the punctured
code threshold $\chi(\tau_0)$ corresponding to $\C_{0}$ at the beginning of the
broadcast stage. Let ${\mathcal F}\subseteq {\mathcal S}$ denote the
\emph{reliable set} of cluster members whose sender-to-helper instantaneous SNR
$\theta_{0,j}>\chi(\tau_0)$. The element of $\mathcal F$ is referred to as the
{\it reliable node}. The punctured code threshold theorem and quasi-static
channel assumption imply that reliable nodes can be guaranteed to decode the
message successfully under ML decoding. Next, each \emph{reliable} node sends
an ACK back to the sender over a fast and error-free feedback channel {\em
without} a need for prior decoding. During this slot, reliable nodes listen and
decode the received signal.

\subsection{Forwarding Stage}

Fig.~\ref{fig:sys-model}.b illustrates the transmission interval corresponding
to slots~$1$ through $M-1$ (here, dark solid circles represent reliable nodes).
Node~$k\in \mathcal F$, re-encodes and partitions received information in the
same manner as done by the sender and, consequently, relays $\C_k$
(corresponding to block~$k$) to the destination in slot~$k$. The sender
transmits the left-over blocks in the remaining transmission slots based on
received ACKs. The signal received at the destination corresponds to an IR
scheme with a fixed number of retransmissions, where a retransmission may
experience a different channel quality.
\begin{example}
Let's assume $M=5$, Node $0$ is the sender, the collaborative cluster $\mathcal
S=\{1,2,3,4\}$, and the reliable set $\mathcal F=\{1,3\}$. Based on the
ACK-based feedback from the cluster, the dynamic schedule is shown in
Table~\ref{tab:schedule}. Since Nodes~$1$ and~$3$ are reliable nodes, they send
$\C_1$ and $\C_3$ at Slots~$1$ and~$3$, respectively. The remaining blocks are
transmitted by Node $0$ at Slots $0$, $2$, and $4$.
\begin{table}[bt]
\caption{\small{Dynamic schedule ($M=5$ and $\mathcal F=\{1,3\}$)}}
\label{tab:schedule}
\begin{center}
\begin{tabular}{| c ||c|c|c|c|c|}
\hline     Slot             & $0$ &   $1$ &  $2$  &  $3$ &  $4$ \\
\hline  Transmitted block   &  $\C_0$ & $\C_1$ & $\C_2$ & $\C_3$ &       $\C_4$       \\
\hline  Transmitter         &  Node $0$ & Node $1$ & Node $0$& Node $3$ &   Node $0$ \\
\hline
\end{tabular}
\end{center}
\end{table}
\end{example}

\begin{remark}
{\it Decoding Delay}: We note that each reliable node needs to decode the
message before it relays this information. This results in a decoding delay in
the transmission. If one of the reliable nodes is scheduled to send the message
in slot $1$, it requires some extra time $\tau_D\cdot T$ between slot $0$ and
slot $1$ due to a decoding latency. In Appendix~\ref{app:rtau}, we propose both
early stopping and a threshold adjusting technique to reduce this time in
practice. For simplicity of the error performance analysis, we assume
$\tau_D=0$ in Sec.~\ref{sec:fer}.
\end{remark}

\begin{remark} {\it Decoding Failure}: Since the assignment rate $\tau_0$ is
predetermined, each helper can pre-calculate the punctured code threshold
$\chi(\tau_0)$ by using (\ref{eq:punctured}).  On the other hand, Node $j$ may
estimate the instantaneous received SNR $\theta_{0,j}$ at the beginning of
slot~$0$. In our proposed scheme, if $\theta_{0,j}>\chi(\tau_0)$, Node $j$ is
called reliable node and sends an ACK message to the sender. Next, if Node $j$
fails in decoding the message, this node will stay silent during the forwarding
stage. We refer to this event as the {\it decoding failure}. However,
Theorem~\ref{th:pcth} implies that the probability of such events approaches
zero as the codeword length $N\rightarrow\infty$. In this paper, we focus on
long codes. Hence, we neglect the decoding failure event in the error
performance analysis.
\end{remark}

\section{IR Cooperative Coding Performance Based on Code Outage} \label{sec:fer}
The decoding is performed at the destination upon completion of $M$
transmission slots. The proposed cooperative coding scheme implies that the
received signal is always the mother codeword $\C$ modified by the fading
channel. Moreover, all communication links experience independent quasi-static
Rayleigh fading channels. Thus, the codeword $\C$ is, equivalently, transmitted
over $M$ slots and experiences $|\mathcal F|+1$ independent channel gains.
Consequently, we study the performance of codes transmitted over a block fading
channel \cite{bigl:proa:sham,rpe-IT-II}.

\subsection{Code Outage for a $Q$-Block Fading Channel}

Here, we consider the block-fading channel model \cite{bigl:proa:sham} with $Q$
fading blocks (a group of $Q$ blocks will be referred to as a \emph{frame}),
where the fading coefficient is essentially invariant during a single block and
different from one block to another. Let $\snr_{j}$ and $\nu_{j}$ be the the
average received SNR and the channel power of block $j$, respectively. The
Bhattacharyya parameter $\gamma_j$ is a function of $\snr_{j}$ and $\nu_{j}$,
i.e.,
\[\gamma_{j}=\exp(-\nu_{j}\cdot \snr_{j}) \quad \text{for} \quad
j=0,\dots,Q-1.\]
Hence, the average Bhattacharyya parameter over $Q$
blocks
$$\bar{\gamma}(\bm{\nu})= \sum_{j=0}^{Q-1}\tau_j \gamma_j=\sum_{j=0}^{Q-1}\tau_j\exp(-\nu_{j}\cdot \snr_{j})$$
is a function of the random vector
$\bm{\nu}\triangleq\{\nu_{0},\dots,\nu_{Q-1}\}$ and, thus, for a given good
code, there is a non-negligible probability that the effective Bhattacharyya
distance $-\ln \bar{\gamma}(\bm{\nu})$ is less than the code threshold $\cth$,
termed \emph{code outage} probability. Thus, the average frame error
probability is a function of both the fading distribution and the threshold of
the code ensemble $[\C]$. More precisely, the average ML decoding word error
probability for a good code ensemble $[\C]$ transmitted over a $Q$-block fading
channel can be bounded as follows:
\begin{align}
\bar{P}_{W}(\bar{\gamma},N) &\triangleq \E \bigl[P_{W}(\bar{\gamma},N)\bigr]
\nonumber\\
&={\rm P}\{{\rm error}(N),\,-\ln \bar{\gamma}(\bm{\nu}) \le \cth\}+
{\rm P}\{{\rm error}(N),\,-\ln \bar{\gamma}(\bm{\nu})>\cth\} \nonumber\\
&\leq {\rm P}\{-\ln \bar{\gamma}(\bm{\nu})\le \cth\}+ {\rm P}\{{\rm
error}(N)\,|-\ln \bar{\gamma}(\bm{\nu})>\cth\} \label{eq:bf0}
\end{align}
where ${\rm error}(N)$ stands for the event of the decoding error for the code
of length $N$. Theorem~\ref{th:cth} implies that the second term of
(\ref{eq:bf0}) approaches zero as the code length $N$ increases. Hence, the
following error rate bound holds
\begin{align}
\bar{P}_{W}(\bar{\gamma},N)&\leq {\rm P}\{ \bar{\gamma}(\bm{\nu})\ge
\exp(-\cth)\}+o(1) \label{eq:fer:bf}
\end{align}
where $o(1)\stackrel{N}{\longrightarrow} 0$. We focus on long codes in the
following analysis. To simplify notation, we will omit $N$ and $o(1)$.

\subsection{IR Cooperative Coding Performance}
Here we study the FER performance of the IR cooperative coding scheme for a
quasi-static frequency-flat Rayleigh-fading channel based on the code outage
upper bound (\ref{eq:fer:bf}).

In slot~$0$, the sender  (Node~$0$) broadcasts its information by sending the
punctured codeword $\C_0$. The channel powers $\nu_{0,j}$, $j=1,\dots,M-1$, are
i.i.d. exponential random variables invariant during each transmission period.
The reliable set ${\mathcal F}$ is now randomly distributed over the {\it
collection} of $2^{M-1}$ subsets of ${\mathcal S}$ with probability
\begin{align}
{\rm P}(\mathcal F)&=\prod_{j\in \mathcal F}{\rm
P}\{\theta_{0,j}>\chi(\tau_0) \}\prod_{j\in {\mathcal F}^{c}}{\rm
P}\{\theta_{0,j}\le \chi(\tau_0)
\}\nonumber\\
&=\prod_{j\in \mathcal
F}\exp\bigl[-\chi(\tau_0)\snr_{0,j}^{-1}\bigr] \prod_{j\in {\mathcal
F}^{c}
}\bigl\{1-\exp\bigl[-\chi(\tau_0)\snr_{0,j}^{-1}\bigr]\bigr\},
\label{eq:cop}
\end{align}
where ${\mathcal F}^{c}\triangleq {\mathcal S} \setminus {\mathcal F}.$

For a given $\mathcal F$, the IR cooperation scheme allows the mother codeword
to be transmitted to the destination (Node~$M$) over $M$ slots with $|\mathcal
F|+1$ independent quasi-static fading gains. Hence, the slot~$i$ Bhattacharyya
parameter is
\begin{align}
\gamma_i=\left\{
\begin{array}{cl}
\exp(-\theta_{i,M}) & i\in\mathcal F, \\
\exp(-\theta_{0,M}) & i\in{\mathcal F}^{c}\cup\{0\}.
\end{array}
\right.
\end{align}
Consequently, the Bhattacharyya parameter averaged over $M$ slots is now
\begin{align}
\bar{\gamma}(\bm{\nu},\mathcal F) &=\bigl(1-\sum_{i\in \mathcal
F}\,\tau_{i}\bigr)\exp(-\theta_{0,M})+
\sum_{i\in \mathcal F}\,\tau_{i}\exp(-\theta_{i,M})\nonumber\\
&=\bigl(1-\sum_{i\in \mathcal F}\,\tau_{i}\bigr)\exp(-\nu_{0,M}\snr_{0,M})+ \sum_{i\in \mathcal
F}\,\tau_{i}\exp(-\nu_{i,M}\snr_{i,M}),
\end{align}
where $\bm{\nu}=\{\nu_{0,M},\nu_{1,M}\dots,\nu_{M-1,M}\}$ is a random vector
$M$-tuple with an independent exponential distribution. The bound
(\ref{eq:fer:bf}) implies that the conditional average word error probability
given a reliable set $\mathcal F$ can be bounded as
\begin{align}
\bar{P}_{W}\bigl(\bar{\gamma}\;|\;{\mathcal F}\bigr) &\leq {\rm
P}\bigl\{\bar{\gamma}(\bm{\nu},\mathcal F)\ge \exp(-\cth)\bigr\}
=\int_{\mathcal A}\,\prod_{i=0}^{M-1}e^{-\nu_{i,M}} \,d
\bm{\nu}\nonumber\\
& \triangleq {\mathcal G}(M,\mathcal F,\bm{\snr})\label{eq:funcG}
\end{align}
where $\mathcal A\triangleq \bigr\{\bm{\nu}:\; \bar{\gamma}(\bm{\nu},\mathcal
F)\ge \exp(-\cth)\bigl\}$, $\bm{\snr}=\{\snr_{0,M},\dots,\snr_{M-1,M}\}$, and
${\mathcal G}(M,\mathcal F,\bm{\snr})$ is referred to as the code outage
probability for a given reliable set $\mathcal F$. The ML decoding FER for
cooperative coding scheme averaged over all possible reliable sets can be
bounded as follows
\begin{align}
\textsf{FER}^{(M)}&=\sum_{\text{all possible}~\mathcal F}{\rm P}(\mathcal
F)\bar{P}_{W}\bigl(\bar{\gamma}\;|\;{\mathcal F}\bigr)\le \sum_{\text{all
possible}~\mathcal F}{\rm P}(\mathcal F)\cdot{\mathcal G}(M,\mathcal
F,\bm{\snr}) \label{eq:tof}
\end{align}
where the superscript $(M)$ represents the number of (potential)
transmitting nodes.

\begin{example}[$M=1$]
The case $M=1$ is equivalent to the traditional direct transmission between the
sender and the destination, and the single-hop FER is
\begin{align}
\textsf{FER}^{(1)}&\le{\mathcal G}(1,\emptyset,\snr_{0,1})\nonumber\\
&={\rm P}\{\nu_{0,1}\snr_{0,1}\le \cth\}\nonumber\\
&=1-\exp(-\cth\snr_{0,1}^{-1}).
\end{align}
\end{example}

\begin{example}[$M=2$]
Here we consider the IR cooperative coding scheme with a single
helper. In this case, the scheme FER is
\begin{align}
\textsf{FER}^{(2)}&\le{\rm P}(\mathcal
F=\emptyset)\bar{P}_{W}\bigl(\bar{\gamma}\;|\; \emptyset \bigr) + {\rm
P}(\mathcal
F=\{1\})\bar{P}_{W}\bigl(\bar{\gamma}\;|\;\{1\}\bigr)\nonumber\\
&= \Bigl[1-e^{-\,\chi(\tau)\snr_{0,1}^{-1}}\Bigl]\,{\mathcal
G}(2,\emptyset,\bm{\snr})+e^{-\,\chi(\tau)\snr_{0,1}^{-1}} \,{\mathcal
G}(2,\{1\},\bm{\snr}), \label{eq:nonfer2-0}
\end{align}
where (assuming $\tau_0,\tau_1\le \exp(-\cth)$)
\begin{align}
{\mathcal G}(2,\emptyset,\bm{\snr})&=1-\exp(-\cth\snr_{0,2}^{-1}) \label{eq:nonfer2}\\
{\mathcal G}(2,\{1\},\bm{\snr})&=1-\omega -\int_{\omega}^{1}
\Bigl[\frac{\exp(-\cth)-\tau_0
x^{\snr_{0,2}}}{\tau_1}\Bigr]^{1/\snr_{1,2}}\,dx, \label{eq:cofer2}
\end{align}
$\omega=\exp\bigl[-\chi(\tau_0)\cdot\snr_{0,2}^{-1}\bigr]$, and the
intermediate steps for deriving (\ref{eq:cofer2}) are given in
Appendix~\ref{app:cof}.
\end{example}
In general, (\ref{eq:tof}) cannot be calculated in a closed form and one needs to resort to
numerical integration methods.

\section{Asymptotic Analysis}\label{sec:limit}

In this section we consider several different cooperation scenarios and derive
asymptotic (in SNR) FER bounds, which have a closed form. For simplicity, we
assume
\begin{align}
\tau_0,\dots,\tau_{M-1}> 1-\exp(-\cth). \label{eq:tauc}
\end{align}
i.e., each randomly punctured code is self-decodable (with probability one).
Next, we refer to $r=\max[d_{0,1},\cdots,d_{0,M-1}]$ as the sender-to-cluster
distance, $D=d_{0,M}$ as the sender-to-destination distance, and
$d=\max[d_{1,M},\cdots,d_{M-1,M}]$ as the cluster-to-destination distance as
shown in Fig~\ref{fig:dis}. Similarly, we define the sender-to-cluster SNR, the
sender-to-destination SNR, and the cluster-to-destination SNR as
\begin{align}
\rho\triangleq E\cdot r^{-L},~~\eta=E\cdot
D^{-L},~~\text{and}~~\lambda\triangleq E\cdot d^{-L}.
\label{eq:SNR-IO}
\end{align}

Computing FER bound (\ref{eq:tof}) requires integration in ${\mathcal
G}(M,\mathcal F,\bm{\snr})$, which presents the code outage probability for a
given $\mathcal F$. The following theorem which is the basis of our asymptotic
analysis, among their contributions, helps in avoiding this integration.
\begin{theorem}\label{th:lim}
Consider $Q$ independent random variables $\phi_{1},\dots,\phi_Q$
with the following properties:
\begin{align*}
0 < \phi_m\le 1 \quad \text{for}~m=1,\dots,Q
\end{align*}
where the probability distribution of $\phi_m$ is a function of $\lambda_m$
such that
\begin{align}
\lim_{\lambda_m\rightarrow\infty}\lambda_m\,{\rm P}\bigl[\phi_m>c\bigr]=-\ln c
\label{eq:u1}
\end{align}
and $0<c<1$. If $\tau_1,\dots,\tau_Q>1- c$ and $\sum_{m=1}^{Q}\tau_m=1,$ then
\begin{align}
\limsup_{\lambda_1,\dots,\lambda_Q\rightarrow\infty}\prod_{m=1}^{Q}\lambda_{m}{\rm
P}\left[\sum_{m=1}^{Q}\tau_m \phi_m > c\right]
\le\frac{1}{Q!}\prod_{m=1}^{Q}\ln\frac{\tau_m}{c-(1-\tau_m)},
\label{eq:lm2-p0}
\end{align}
where $\lambda_1,\dots,\lambda_Q\rightarrow\infty$ means
$\lambda_1\rightarrow\infty,\dots,\lambda_Q\rightarrow\infty.$

\begin{proof}
The proof is provided in Appendix~\ref{app:bound} based on the induction
method.
\end{proof}
\end{theorem}

\subsection{Transmitter Clustering}

In the {\it transmitter clustering} scenario we assume that $M-1$ helpers are
very close to the sender so that $r\rightarrow 0$. For this setting, we call
the sender-to-cluster channel {\it fully reliable} in the sense of ${\rm
P}({\mathcal F}=\{1,\dots,M-1\})= 1,$ i.e., all helpers are reliable nodes with
probability one. Thus, the cooperation scheme FER can be written as
\begin{align}
\textsf{FER}^{(M)}_{\rm T}
&=\bar{P}_{W}\bigl(\bar{\gamma}\;|\;{\mathcal
F}=\{1,\dots,M-1\}\bigr)\nonumber\\
&\le {\rm P}\{-\ln \bar{\gamma}(\bm{\nu},\{1,\dots,M-1\})\le \cth\} \nonumber\\
&= {\mathcal G}(M,\mathcal
F=\{1,\dots,M-1\},\bm{\snr}=\{Ed_{i,M}^{-L}\}). \label{eq:trc}
\end{align}
Now, let's consider the large SNR case. Note that the exponential distribution
of $\nu_{i,M}$ implies that
\begin{align}
&\lim_{\snr_{i,M}\rightarrow \infty}\, \snr_{i,M}\,{\rm P}\Bigl[\exp(-\nu_{i,M}\snr_{i,M}) \geq
\exp(-\cth)\Bigr]\nonumber\\
=&\lim_{\snr_{i,M}\rightarrow \infty}\,\frac{1-\exp(-\cth\snr_{i,M}^{-1})}{\snr_{i,M}^{-1}}\nonumber\\
=&\cth.\label{eq:con1}
\end{align}
Thus, (\ref{eq:punctured}), (\ref{eq:funcG}), (\ref{eq:tauc}), and  Theorem~\ref{th:lim} imply
\begin{align}
\limsup_{\boldsymbol{\snr}\rightarrow \infty}\,
\prod_{i=0}^{M-1}\snr_{i,M}\cdot {\mathcal
G}(M,\{1,\dots,M-1\},\bm{\snr}) \le
\frac{1}{M!}\prod_{i=0}^{M-1}\chi(\tau_i).
\end{align}
For large enough $E$, we can rewrite (\ref{eq:trc}) as
\begin{align}
\textsf{FER}^{(M)}_{\rm T}&\le_{E}\frac{1}{M!}\prod_{i=0}^{M-1} \chi(\tau_i)\snr_{i,M}^{-1}\label{eq:lsnr-t}\\
&=\frac{1}{E^{M} M!}\prod_{i=0}^{M-1}\chi(\tau_i)d_{i,M}^{L} \\
&=\frac{D^{ML}}{E^{M}
M!}\prod_{i=0}^{M-1}\chi(\tau_i)\label{eq:limit-t}
\end{align}
where $\le_{E}$ means that the inequality holds for  sufficiently
large $E$, and the last step is based on the triangle inequality
$D-r\le d_{i,M}\le D+r$ and $r\rightarrow 0$.

\subsection{Receiver Clustering} \label{subb:cp}
In the {\it receiver clustering} scenario we assume that $M-1$ cluster members
are very close to the destination so that $d\rightarrow 0$. Note that
(\ref{eq:tauc}) implies that each block (punctured codeword) is self-decodable
with probability one. Hence, the code outage probability is zero for any
nonempty reliable set $\mathcal F$, i.e,
\begin{align}
{\mathcal G}(M,\mathcal F,\bm{\snr})=0~~\text{for}~\mathcal F\neq\emptyset.
\end{align}
Therefore, we can bound the cooperation scheme FER as
\begin{align}
\textsf{FER}^{(M)}_{\rm R}&={\rm P}({\mathcal
F}=\emptyset)\bar{P}_{W}^{[\mathcal
C]}\bigl(\bar{\gamma}\;|\;{\mathcal
F}=\emptyset\bigr)\nonumber\\
&\le {\rm P}({\mathcal F}=\emptyset){\mathcal G}(M,\mathcal F=\emptyset,\bm{\snr}) \nonumber\\
&=
\prod_{j=1}^{M-1}\bigl\{1-\exp\bigl[-\chi(\tau_0)\snr_{0,j}^{-1}\bigr]\bigr\}
\bigl[1-\exp(-\cth\eta^{-1})\bigr] \nonumber\\
&=\prod_{j=1}^{M-1}\bigl\{1-\exp\bigl[-\chi(\tau_0)E^{-1}d_{0,j}^{L}\bigr]\bigr\}
\bigl[1-\exp(-\cth E^{-1}D^{L})\bigr]\label{eq:rec}
\end{align}
Again, we focus on the large SNR case. Note that
\begin{align*}
\lim_{\snr\rightarrow
\infty}\snr\cdot\bigl[1-\exp(-a\snr^{-1})\bigr]=a \quad
\text{for}~a>0.
\end{align*}
Hence, for large enough $E$, we can rewrite (\ref{eq:rec}) as
\begin{align}
\textsf{FER}^{(M)}_{\rm R}&\le_{E}
\bigl[\chi(\tau_0)\bigr]^{M-1}\cth
\cdot\prod_{j=1}^{M}\snr_{0,j}^{-1} \label{eq:lsnr-r}\\
&=\frac{\bigl[\chi(\tau_0)\bigr]^{M-1}\cth}{E^{M}}
\cdot\prod_{j=1}^{M}d_{0,j}^{L} \nonumber \\
&=\frac{\bigl[\chi(\tau_0)\bigr]^{M-1}\cth}{E^{M}} \cdot D^{ML}
\label{eq:limit-r}
\end{align}
where the last step follows from the geometric property $D-d\le d_{0,j}\le D+d$
and $d\rightarrow 0$.

\subsection{Cluster Hopping}
\begin{figure}[hbt]
  \centerline{\includegraphics[width=0.6\linewidth,draft=false]{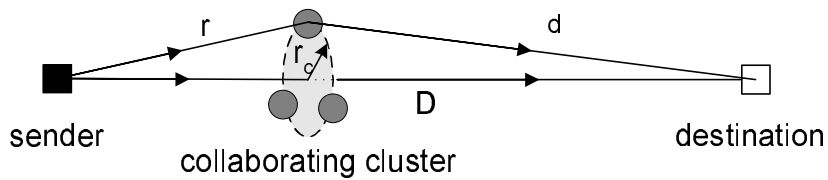}}
  \caption{\small Cluster hopping} \label{fig:exp}
\end{figure}

Here, we assume that $d_{0,1}=\dots=d_{0,M-1}=r>0$ and
$d_{1,M}=\dots=d_{M-1,M}=d>0$. The distances $d$, $r$, and $D$
satisfy the triangle inequality as shown in Fig.~\ref{fig:exp}. By
using (\ref{eq:cop}) in the high SNR regime, we have
\begin{align}
\lim_{\boldsymbol{\snr}\rightarrow \infty}\prod_{j\in {\mathcal
F}^c\setminus \{0\}}\snr_{0,j}\,{\rm P}(\mathcal F)
&=\lim_{\rho\rightarrow \infty} \rho^{M-(|\mathcal F|+1)}\,{\rm P}(\mathcal F)\nonumber\\
&=\lim_{\rho\rightarrow \infty}
\Biggl\{\exp\bigl[-\chi(\tau_0)\rho^{-1}\bigr]\Biggr\}^{|\mathcal
F|}\Biggl\{\frac{1-\exp(-\chi(\tau_0)\rho^{-1})}{\rho^{-1}}\Biggr\}^{M-(|\mathcal F|+1)}\nonumber\\
&=[\chi(\tau_0)]^{M-(|\mathcal F|+1)}.
\end{align}
Moreover, (\ref{eq:punctured}), (\ref{eq:funcG}), (\ref{eq:tauc}),
(\ref{eq:con1}), and Theorem~\ref{th:lim} imply
\begin{align}
\limsup_{\boldsymbol{\snr}\rightarrow \infty}\, \prod_{i\in
{\mathcal F} \cup \{0\}}\snr_{i,M}\cdot {\mathcal G}(M,\mathcal
F,\bm{\snr})
&= \limsup_{\lambda,\,\eta \rightarrow \infty}
(\lambda^{|\mathcal F|}\,\eta)\cdot {\mathcal
G}(M,\mathcal F,\bm{\snr})  \nonumber\\
&\le\frac{\chi(1-\sum_{i\in {\mathcal F}}\tau_i)}{(|\mathcal
F|+1)!}\prod_{i\in {\mathcal F}}\chi(\tau_i).
\end{align}
Note that
\begin{align*}
\snr_{0,M}=\eta=ED^{-L},\quad \snr_{0,j}&=\rho=Er^{-L},\quad
\snr_{j,M}=\lambda=Ed^{-L},\quad j=1,\dots,M-1.
\end{align*}
Hence, for large enough $E$, we can rewrite (\ref{eq:tof}) as
\begin{align}
\textsf{FER}^{(M)}&\le_{E} \sum_{\mathcal F}
\Biggl\{\frac{[\chi(\tau_0)]^{M-(|\mathcal F|+1)} \chi(1-\sum_{i\in {\mathcal
F}}\tau_i)}{(|\mathcal F|+1)!} \prod_{i\in {\mathcal F}}\chi(\tau_i)\Biggr\}
(\rho^{|\mathcal F|+1-M} \lambda^{-|\mathcal F|}\,\eta^{-1}) \label{eq:lsnr-c}\\
&= \sum_{\mathcal F}
\underbrace{\Biggl\{\frac{[\chi(\tau_0)]^{M-(|\mathcal F|+1)}
\chi(1-\sum_{i\in {\mathcal F}}\tau_i)}{(|\mathcal F|+1)!}
\prod_{i\in {\mathcal F}}\chi(\tau_i)\Biggr\}}_{\text{coding
advantage}} \cdot \underbrace{(r^{M-|\mathcal F|-1} d^{|\mathcal
F|}\,D)^{L}}_{\text{geometric distance profile}}\cdot E^{-M}.
\label{eq:limit2}
\end{align}

\subsection{Diversity Gain}
Following \cite{zheng:tse}, the \emph{diversity gain} is defined as
\begin{align}
\textsf{div}\triangleq \lim_{\snr \rightarrow \infty}\frac{-\log \textsf{FER}}{\log \snr}.
\end{align}
Since our collaborating model is a distributed {\it multiple-input
single-output} (MISO) system, the maximum achievable diversity gain is $M$.
Equations (\ref{eq:lsnr-t}), (\ref{eq:lsnr-r}), and (\ref{eq:lsnr-c})
illustrate that all of the three discussed scenarios: transmitter clustering,
receiver clustering, and cluster hopping can achieve the {\it full} diversity
gain, i.e., $\textsf{div}=M$, in high SNR regime.

\subsection{Cooperative Coding Gain}

For small $\cth$, we can build the following simple relationship between the punctured code
threshold and the mother code threshold $\cth$. Equation (\ref{eq:punctured}) implies
\begin{align}
\lim_{\cth\rightarrow 0}\frac{\chi(\tau)}{\cth}&=\lim_{\cth\rightarrow 0}
(\cth)^{-1}\ln \frac{\tau}{\exp(-\cth)-(1-\tau)}=\frac{1}{\tau}.
\end{align}
Thus, we can rewrite (\ref{eq:limit2}) as
\begin{align}
\textsf{FER}^{(M)}\le_{E,\cth}\sum_{\mathcal F}
\Biggl\{\frac{\bigl(\cth\bigr)^{M}} {\tau_0^{M-(|\mathcal F|+1)}
(1-\sum_{i\in {\mathcal F}}\tau_i) (\prod_{i\in {\mathcal F}}\tau_i)
(|\mathcal F|+1)!} \Biggr\} \cdot (r^{M-|\mathcal F|-1} d^{|\mathcal
F|}\,D)^{L}\cdot E^{-M}.
 \label{eq:limit3}
\end{align}
where $\le_{E,\,\cth}$ means that the inequality holds for sufficiently large $E$ and small
$\cth$.
\begin{example}[$M=1$ limiting case] \label{ex:1-limit}
\begin{align}
\textsf{FER}^{(1)}\le_{E}\, \frac{\cth}{\eta}=\frac{\cth}{E} D^{L} .
\label{eq:1-app}
\end{align}
\end{example}
\begin{example}[$M=2$ limiting case]
\begin{align}
\textsf{FER}^{(2)}\le_{E,\,\cth}\,\frac{\bigl(\cth\bigr)^2}{\tau_0\,\eta\,\rho}+
\frac{\bigl(\cth\bigr)^2}{2\tau_0\tau_1\,\eta\,\lambda}=\left(\frac{\cth}{E}\right)^2\left[\frac{(r\,
D)^L}{\tau_0}+ \frac{(d\,D)^L}{2\tau_0\tau_1}\right].
\label{eq:2-app}
\end{align}
\end{example}
Similarly, (\ref{eq:limit-t}) and (\ref{eq:limit-r}) can be rewritten as
\begin{align}
\textsf{FER}^{(M)}_{\rm T} &\le_{E,\,\cth} \frac{1}{M!\prod_{i=0}^{M-1}\tau_i }
\left(\frac{\cth D^{L}}{E}\right)^M\label{eq:limt-c}\\
\textsf{FER}^{(M)}_{\rm R}&\le_{E,\,\cth}
\tau_0^{-(M-1)}\left(\frac{\cth D^{L}}{E}\right)^M.
\label{eq:limt-r}
\end{align}
In \cite{lane2}, the author defines the \emph{cooperative coding
gain} as
\begin{align}
\textsf{cop}\triangleq \lim_{\snr \rightarrow \infty}\frac{\textsf{FER}^{-1/\textsf{div}}}{\snr}.
\end{align}
Let the sender-to-destination SNR be the basis, i.e, $\eta=\snr$. Bounds
(\ref{eq:limt-c}), (\ref{eq:limt-r}), and (\ref{eq:limit3}) imply that the
coding gains of transmitter clustering, receiver clustering, and cluster
hopping schemes satisfy
\begin{align}
\lim_{\cth\rightarrow 0} \cth \cdot \textsf{cop}^{(M)}_{\rm T}&\ge
\left(M!\prod_{i=0}^{M-1}\tau_i \right)^{1/M},\\
\lim_{\cth\rightarrow 0} \cth \cdot \textsf{cop}^{(M)}_{\rm R}&\ge \tau_0^{(M-1)/M},\\
\lim_{\cth\rightarrow 0} \cth \cdot \textsf{cop}^{(M)} &\ge
\left\{\sum_{\mathcal F} \frac{\bigl[(r/D)^{M-|\mathcal F|-1}
(d/D)^{|\mathcal F|}\bigr]^{L}}{\tau_0^{M-(|\mathcal F|+1)}
(1-\sum_{i\in {\mathcal F}}\tau_i) (\prod_{i\in {\mathcal F}}\tau_i)
(|\mathcal F|+1)!}\right\}^{-1/M}. \label{eq:ccod}
\end{align}
Inequality (\ref{eq:ccod}) illustrates that, in general, the cooperative coding
gain is a function of the code parameter $\cth$, the cooperation scheme
parameters $\{\tau_i\}$,  and the geometric distance profile $(r,\,d,\,D)$ of
the network.

\section{Simulations and Discussions}\label{sec:sd}

\subsection{IR Cooperative Turbo Coding}

In this subsection we study the error performance of the IR cooperation scheme
based on the turbo code described in Example~\ref{ex:tc-cth}. FER simulations
consider binary antipodal signaling and an independent flat quasi-static
Rayleigh fading for each link.
\begin{figure}[bt]
  \centerline{\includegraphics[width=0.65\linewidth,draft=false]{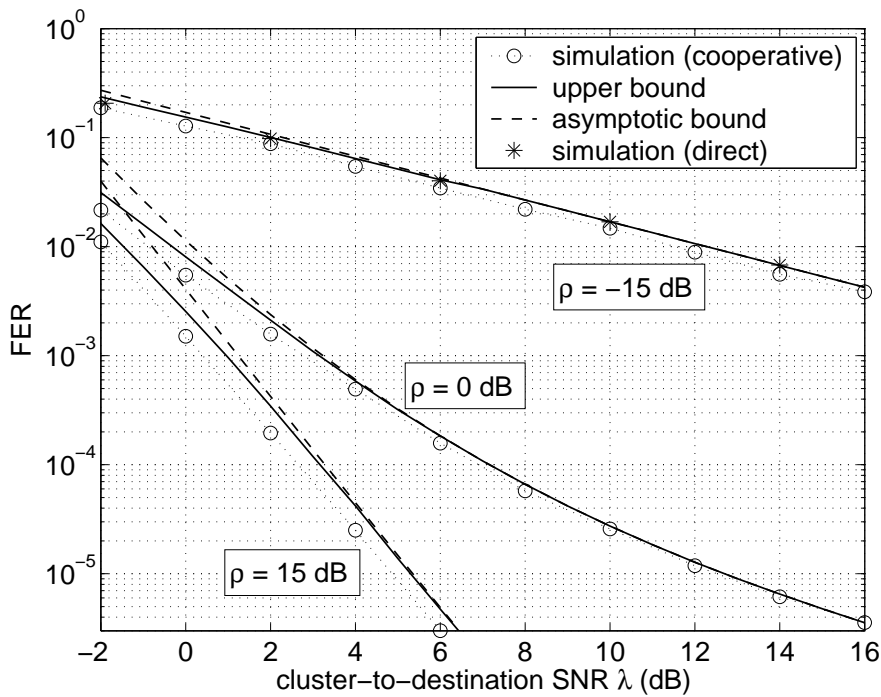}}
  \caption{\small  FER vs. $\lambda$
  ($M=5$, $\rho=-15,\,0,\,15$, and mother turbo code of rate $R=1/7$ and length $N=5376$)} \label{fig:cd}
\end{figure}
Each receiver has perfect channel state information and employs coherent
detection. All receivers employ a multiple turbo decoder based on the triangle
iterative decoding algorithm \cite{divs:poll-1}.

\begin{figure}[hbt]
  \centerline{\includegraphics[width=0.65\linewidth,draft=false]{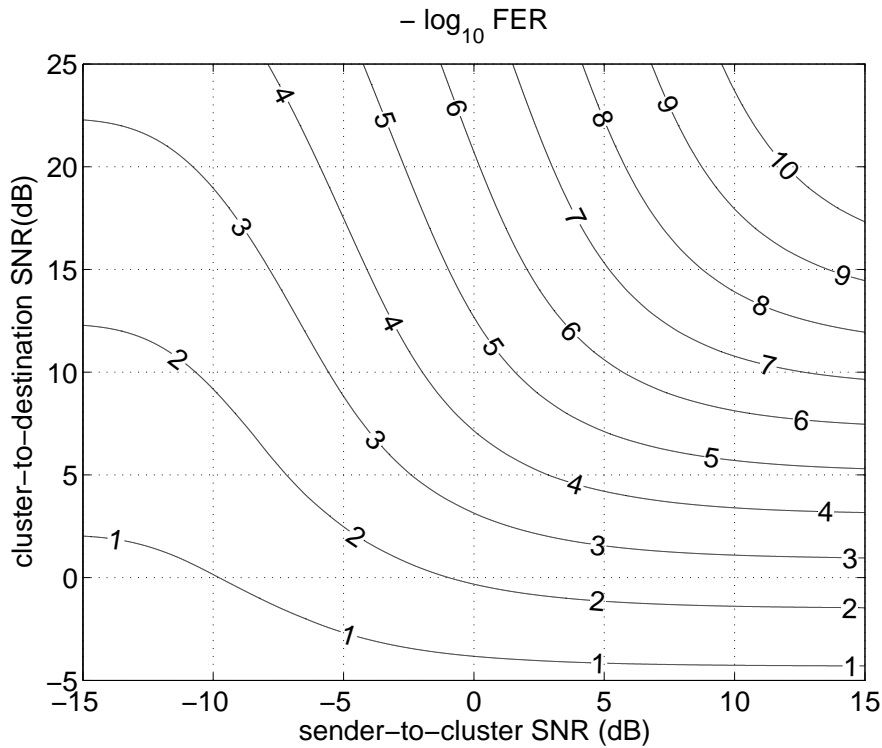}}
  \caption{\small  FER performance as a function of $\rho$ and $\lambda$
  ($M=5$ and mother turbo code of rate $R=1/7$)} \label{fig:IO}
\end{figure}

Here we consider a $M=5$ collaborative network and assume
$\snr_{0,1}=\dots=\snr_{0,4}=\rho$ and
$\eta=\snr_{0,5}=\dots=\snr_{4,5}=\lambda.$ Thus, the FER performance of
cooperative turbo codes is a function of both cluster-to-destination SNR $\rho$
and sender-to-cluster SNR $\lambda$. Fig.~\ref{fig:cd} depicts the FER for
$\rho=-15,\,0,\,15$ dB and as a function of $\lambda$ from $-2$ to $16$ dB.
\begin{figure}[bt]
  \centerline{\includegraphics[width=0.65\linewidth,draft=false]{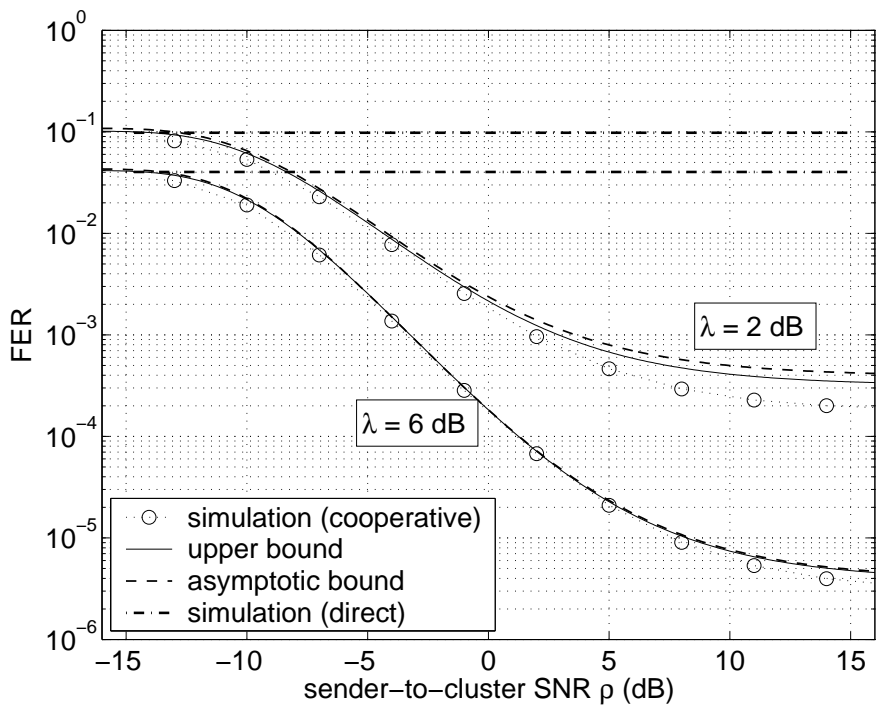}}
  \caption{\small  FER performance vs. $\rho$
  ($M=5$, $\lambda=2,\,6$, and mother turbo code of rate $R=1/7$ and length $N=5376$)}
  \label{fig:ic}
\end{figure}
On the other hand, in Fig.~\ref{fig:ic}, we fix $\lambda=2,6$ dB and study the
FER performance vs. sender-to-cluster SNR $\rho$. For these two cases, we
compare the simulation result with the analytic upper bound (\ref{eq:tof}) and
the asymptotic bound (\ref{eq:limit3}). Figs.~\ref{fig:cd} and~\ref{fig:ic}
also depict the simulation result of the direct transmission as a benchmark. We
observe that the upper bound (\ref{eq:tof}) accurately predicts the cooperative
coding performance and the asymptotic bound (\ref{eq:limit3}) converges to the
bound (\ref{eq:tof}) for medium and high SNR. This observation enables us to
estimate the FER performance as a function of $\rho$ and $\lambda$ by combining
(\ref{eq:tof}) and (\ref{eq:limit3}) in Fig.~\ref{fig:IO}, where we use the
bound (\ref{eq:tof}) for low SNR and employ the bound (\ref{eq:limit3}) to
simplify the computation for medium and high SNR.

\subsection{Collaborative Cluster Size}\label{subsec:M}

Here, we study the effect of the collaborative cluster size on the FER
performance of the transmitter clustering (the sender-to-cluster distance
$r\rightarrow 0$ in this scenario). We assume that nodes have limited battery
energy. In this case, achieving high transmission energy efficiency is more
important than maximizing the diversity gain. Our approach is to assume that
the allowable FER is $\epsilon$, which guarantees the \emph{quality of service}
(QoS), and to determine the $\epsilon$-achievable transmission energy by
applying the asymptotic bound studied in Sec.~\ref{sec:limit}. The closed form
bound predicts well the error performance for medium and high SNR.

Let $\tau_0=\dots=\tau_{M-1}=1/M$, now, (\ref{eq:limt-c}) implies
\begin{align}
\textsf{FER}^{(M)}_{\rm T} &\le_{E,\,\cth}\frac{1}{M!}\left(\frac{M
\cdot \cth D^{L}}{E^{(M)}}\right)^{M}
\end{align}
where $E^{(M)}$ is the $\epsilon$-achievable transmission energy. To satisfy
the QoS requirement, we require\footnote{Strictly speaking, the bounds
(\ref{eq:limit3}), (\ref{eq:limt-c}), and (\ref{eq:limt-r}) are based on the
large SNR assumption. However, through simulations, we observe that the
asymptotic bounds also works well for the medium SNR. On the other hand, these
asymptotic bounds can be expressed in a closed form, whereas, the calculation
of the bound (\ref{eq:tof}) requires numerical integration method. Thus, here
and hereafter, we use these asymptotic bound to estimate the FER performance.}
\begin{align}
\frac{1}{M!}\left(\frac{M \cdot \cth D^{L}}{E^{(M)}}\right)^{M}
=\epsilon.
\end{align}
Based on Stirling's approximation, the $\epsilon$-achievable transmission
energy is
\begin{align}
E^{(M)} \approx \frac{\cth\cdot e D^{L}}{(\epsilon \sqrt{2\pi
M})^{1/M}} .\label{eq:engm}
\end{align}
To illustrate how much energy can be saved using the IR cooperative
transmission, we consider both direct transmission and transmission over a
fully interleaved Rayleigh fading channel cases as benchmarks. For direct
transmission (i.e, $M=1$), by using the bound (\ref{eq:1-app}) in
Example~\ref{ex:1-limit}, the $\epsilon$-achievable energy is given by
\begin{align}
E^{(1)} = \frac{\cth\cdot D^{L}}{\epsilon}  .\label{eq:eng1}
\end{align}
For a fully interleaved Rayleigh fading channel, the Bhattacharyya parameter is
a function of the sender-to-destination SNR $\eta$ (see \cite{huj:mceli} for
the detail) as follows
\begin{align}
\gamma^{({\rm FIRF})}=\frac{1}{1+\eta}
\end{align}
where $({\rm FIRF})$ stands for fully interleaved Rayleigh fading.
Theorem~\ref{th:cth} implies that the asymptotic word error rate
approaches zero (as $N\rightarrow \infty$) if $\gamma^{({\rm
FIRF})}<\exp(-\cth)$, i.e., the sender-to-destination SNR
\[\eta > \exp(\cth)-1.\]
Note that $\eta=E\cdot D^{-L}$. Thus, for this channel, the reliable
transmission energy threshold is defined by
\begin{align}
E^{({\rm FIRF})} \triangleq [\exp(\cth)-1]\cdot D^{L}.
\label{eq:fer-I}
\end{align}

Let $U^{(M)}\triangleq E^{(1)}/E^{(M)}$ denote the transmission {\it
energy saving}. Equations (\ref{eq:engm}) and (\ref{eq:eng1}) lead
to
\begin{align}
U^{(M)}\approx \frac{(2\pi M)^{1/2M}}{e \cdot \epsilon^{1-1/M}}
\label{eq:save}
\end{align}
which illustrates the fact that the energy saving $U^{(M)}$ is a
function of only $M$ and $\epsilon$ for the transmitter clustering
scenario and does not depend on the code threshold $\cth$ and the
sender-to-destination distance $D$. In other words, the energy
saving of the IR cooperative coding scheme is {\it universal} for
all good code families and sender-to-destination distances.
\begin{figure}[hbt]
  \centerline{\includegraphics[width=0.65\linewidth,draft=false]{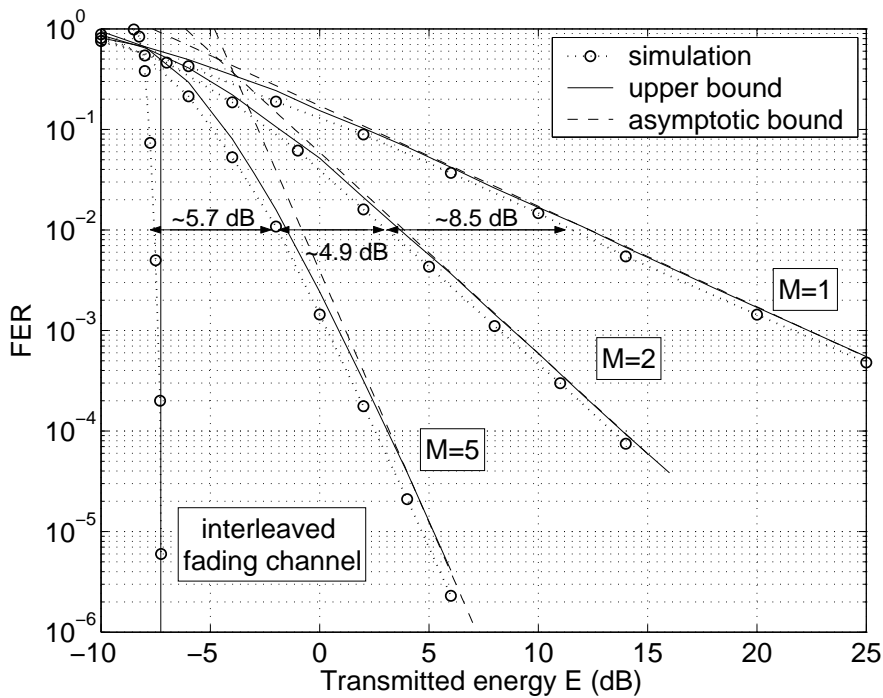}}
  \caption{\small FER vs. transmission energy in the transmitter clustering scenario
  ($D$=1, ${\rm P}(|\mathcal F|=M-1)=1$, and mother turbo code of rate $R=1/7$ and length $N=5376$)} \label{fig:perf}
\end{figure}
\begin{table}
\caption{Energy saving vs. $M$} \label{tab:save}
\begin{center}
\begin{tabular}{| c ||cccc|c|}
\hline          $M$          & $2$ & $3$  & $4$  & $5$  &     ${\rm FIRF}$         \\
\hline  Estimated $U^{(M)}$ (dB)    &  $8.4$ & $11.1$ & $12.4$ & $13.2$ &       $<20$       \\
\hline
\end{tabular}
\end{center}
\end{table}
Next, we set $\epsilon=0.01$ and numerically compute the estimated energy
saving in Table~\ref{tab:save} based on the approximation (\ref{eq:save}). We
compare the calculation result with the fully interleaved fading channel
savings
\begin{align}
U^{({\rm FIRF})}&=\frac{E^{(1)}}{E^{({\rm
FIRF})}}=\frac{\cth}{[\exp(\cth)-1]\cdot\epsilon}<1/\epsilon,
\end{align}
where the inequality follows from $\exp(x)>1+x$ for $x>0$.
Fig.~\ref{fig:perf} illustrates the simulated FER performance versus
transmission energy $E$ as well as the upper bound~(\ref{eq:tof})
and its asymptotic version~(\ref{eq:limit3}) for $D=1$ and
$r\rightarrow 0$. In Fig.~\ref{fig:perf}, we also compare the FER
performance of cooperative transmissions vs. transmission over a
fully interleaved Rayleigh fading channel. In the latter case, the
error performance exhibits a threshold behavior and the reliable
transmission energy is described in (\ref{eq:fer-I}). We observe
that the energy saving obtained through simulation (in
Fig.~\ref{fig:perf}) and the estimated $U^{(M)}$ (in
Table~\ref{tab:save}) illustrate an excellent match. Furthermore,
both Table~\ref{tab:save} and Fig.~\ref{fig:perf} illustrate the
fact that, although the cumulative energy saving increases with $M$,
the rate of increase drops quickly.

\subsection{Normalized Cluster to Destination Distance}

Here, we assume $d_{0,1},\dots,d_{0,M-1}=r$, $d_{1,M}=\dots=d_{M-1,M}=d$,
$D\thickapprox r+d$ and $\tau_1=\dots=\tau_{M-1}$. We move the collaborative
cluster from the sender towards the destination, and evaluate the energy saving
in terms of the normalized cluster to destination distance $\kappa\triangleq
r/D$.
\begin{figure}[hbt]
  \begin{minipage}{.5\linewidth}
 \centerline{\includegraphics[width=\linewidth,draft=false]{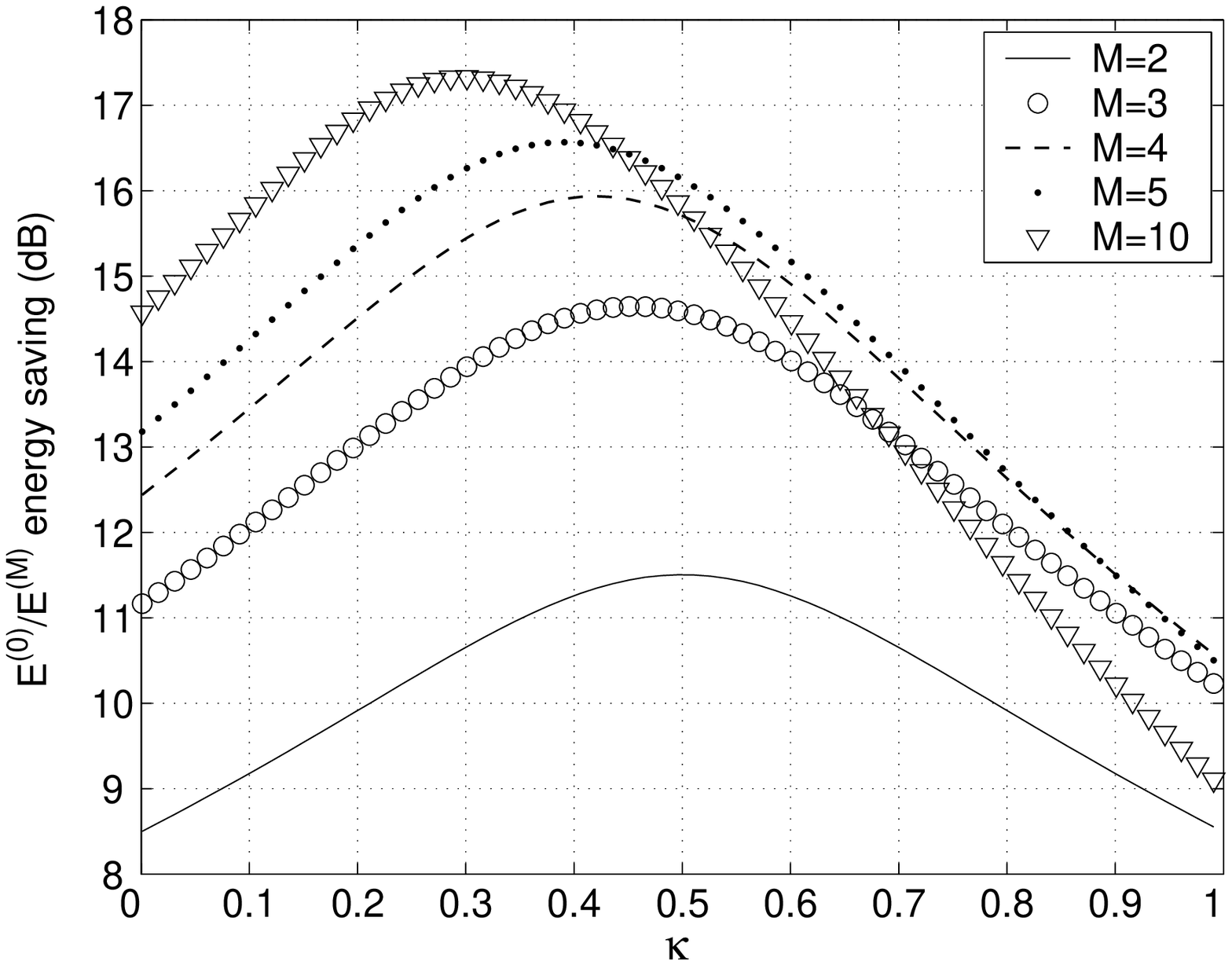}}
   \vspace{-0.1cm}
    \centerline{\mbox{\small a. $\tau_0=1/M$ }}
  \end{minipage}\hfill
  \begin{minipage}{0.5\linewidth}
  \centerline{\includegraphics[width=\linewidth,draft=false]{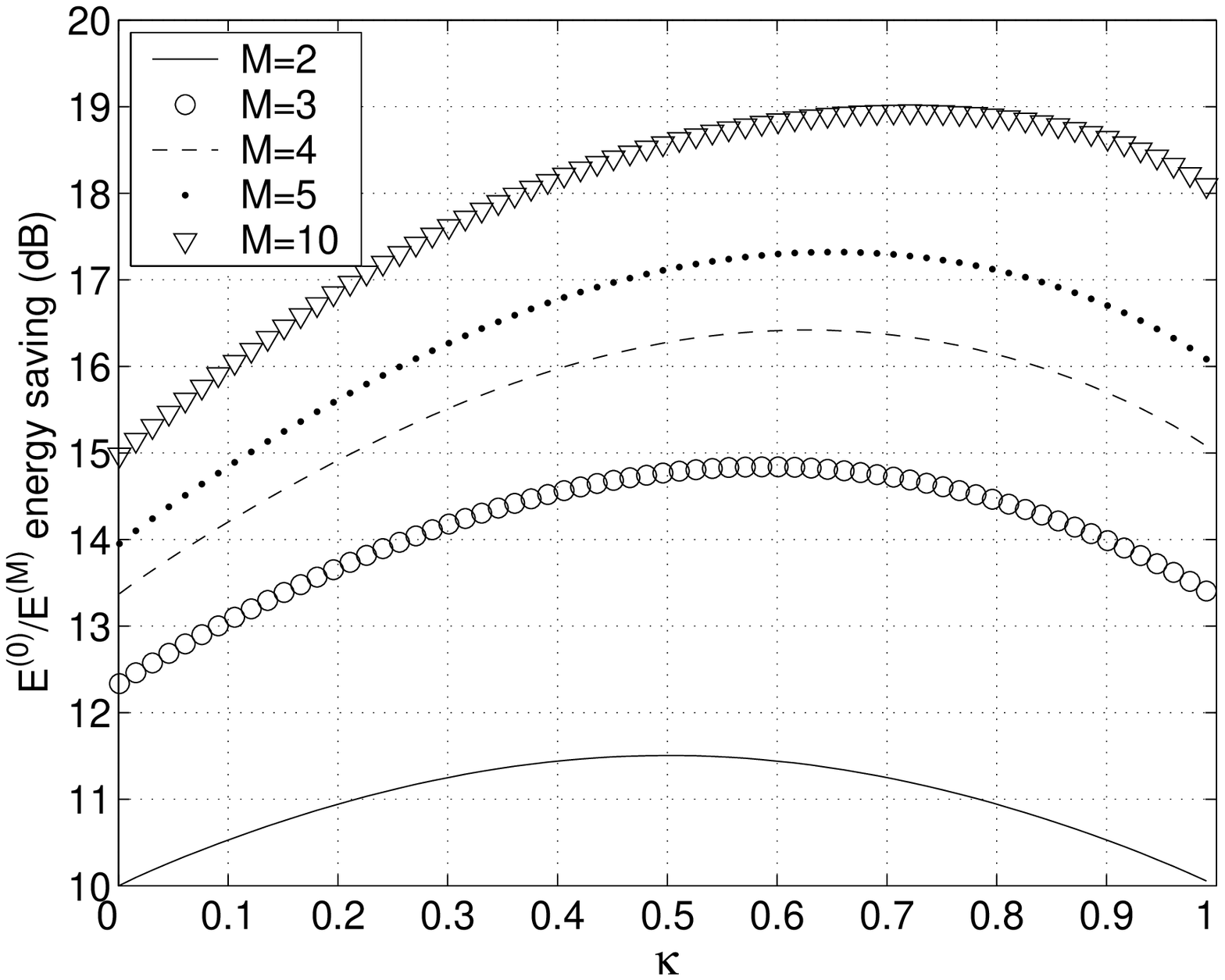}}
    \vspace{-0.1cm}
    \centerline{\mbox{\small b. optimum $\tau_0$}}
  \end{minipage}
  \vspace{-0.2cm}
  \caption{\small $\epsilon$-achievable transmission energy saving vs. normalized cluster distance $\kappa$
  (the required FER $\epsilon=0.01$, path loss exponent
   $L=3$, $\tau_j=(1-\tau_0)/(M-1)$ for $j\neq0$)} \label{fig:energy}
\end{figure}
Here, we use the similar approach as one used in the pervious subsection. We
assume that the allowable FER is $\epsilon$ and calculate the the
$\epsilon$-achievable energy $E^{(M)}$ based on the asymptotic FER bound
(\ref{eq:limit3}), i.e.,
\begin{align}
\sum_{\mathcal F} \Biggl\{\frac{\bigl(\cth\bigr)^{M}(r^{M-|\mathcal F|-1}
d^{|\mathcal F|}\,D)^{L}\cdot \bigl[E^{(M)}\bigr]^{-M}} {\tau_0^{M-(|\mathcal
F|+1)} (1-\sum_{i\in {\mathcal F}}\tau_i) (\prod_{i\in {\mathcal F}}\tau_i)
(|\mathcal F|+1)!} \Biggr\} &=\epsilon
\end{align}
Hence, we have
\begin{align}
E^{(M)}=\cth\cdot D^{L}\Biggl\{\frac{1}{\epsilon}\sum_{k=0}^{M-1}
\binom{M-1}{k} \frac{\kappa^{L(M-k-1)}
(1-\kappa)^{Lk}}{\tau_{0}^{M-(k+1)}\tau_{\dag}^{k}(1-k\tau_{\dag})(k+1)!}\Biggr\}^{1/M}
\end{align}
where $\tau_{\dag}=(1-\tau_0)/(M-1)$. Next, the transmission energy
saving $U^{(M)}$ is given by
\begin{align}
U^{(M)}=\frac{E^{(1)}}{E^{(M)}}=\left\{\epsilon^{M-1} \sum_{k=0}^{M-1}
\binom{M-1}{k} \frac{\kappa^{L(M-k-1)} (1-\kappa)^{Lk}
}{\tau_{0}^{M-(k+1)}\tau_{\dag}^{k}(1-k\tau_{\dag})(k+1)!}\right\}^{-1/M}.
\label{eq:umsa}
\end{align}
Equation (\ref{eq:umsa}) illustrates that $U^{(M)}$ does not depend on the code
threshold $\cth$ and the sender-to-destination distance $D$ for a given
$\kappa$. In this sense, we claim that the transmission energy saving of IR
cooperative coding schemes is universal.

Based on (\ref{eq:umsa}), we depict the transmission energy saving $U^{(M)}$ in
Fig.~\ref{fig:energy} as a function of the normalized cluster distance $\kappa$
for $M=2,$ $3,$ $4,$ $5,$ $10$, where $\epsilon=0.01$, path loss exponent
$L=3$, and $0<\kappa <1$. Here, we consider two cases: fixing $\tau_0=1/M$ and
choosing optimum $\tau_0$ for a given $\kappa$. In Fig.~\ref{fig:energy}.a, we
observe that $U^{(10)}$ is below $U^{(5)}$ when $\kappa\rightarrow 1$. The
reason is that here the assignment rate is fixed to $\tau_0=1/M$. We note that
$\kappa\rightarrow 1$ implies that the collaborative cluster is close to the
destination and far from the sender. In this case, the transmission in slot $0$
(i.e., broadcast stage) is more important (see the receiver clustering case in
Sec.~\ref{subb:cp} and the asymptotic bound (\ref{eq:limt-r})), and larger
$\tau_0$ helps in increasing the expected number of reliable nodes which relay
the message to the destination at the forwarding stage. Hence, we obtain a
better energy saving performance for $\tau_0=1/4$ than for $\tau_0=1/10$ when
$\kappa\rightarrow 1$.  As a comparison, we depict the energy saving by
optimizing $\tau_0$ for a given $\kappa$ in Fig.~\ref{fig:energy}.b.

\section{Conclusion}\label{sec:con}

In this paper, we study a cooperative coding scheme for multi-hop wireless
networks with incremental redundancy. Threshold analysis of good code ensembles
enables the analysis of the IR cooperation coding scheme.

First, we study the threshold behavior of good code ensembles for BISOM
Channels. A general relationship among the channel capacity, the Bhattacharyya
rate, and cutoff rate is established for BISOM channels. Based on this
relationship and the modified Shulman-Feder reliable channel region
\cite{rpe-IT-II}, a simple code threshold is proposed. The reliable
communication constraint based on the simple code threshold has the same form
as the UB code threshold introduced in \cite{huj:mceli}, but the former result
is tighter by almost $1~ dB$ for an example of a rate $1/7$ turbo code on an
AWGN channel.

Next, we consider the fading channel. Based on the code outage concept, a
simple threshold bound on FER and its asymptotic version are presented for the
IR cooperative coding scheme. The asymptotic bound is a closed form function of
the simple code threshold, average channel SNR, geometric distances, and the
size of the collaborative network. It enables analysis of the diversity and
coding gains for three different scenarios: transmitter clustering, receiver
clustering, and cluster hopping.

Finally, we simulate the IR cooperative turbo code performance based on the
iterative decoding. Remarkably, our analytical FER bound accurately predicts
the simulated cooperative coding performance at any SNR and the asymptotic
bound agrees very well with the simulation results at medium and high SNRs. We
further discuss the transmission energy gain in terms of collaborative network
size and normalized distance for the transmitter clustering and cluster hopping
scenarios, respectively. In both cases, we observed that the energy saving does
not depend on the sender-to-destination distance and the code threshold. In
this sense, we claim that the energy saving of IR cooperative coding scheme is
universal for all good code families and all initial non-cooperative
hop-distance selections.

\appendix
\subsection{Modified Shulman-Feder Reliable Channel
Region}\label{app:msf}

In this appendix, we state for reference the Modified Shulman-Feder (MSF)
reliable channel region for good binary code ensembles transmitted over a set
of $Q$ (independent) parallel channels (see \cite{rpe-IT-II} for the detail).

\begin{theorem}[cf.~{\cite[Theorem~5]{rpe-IT-II}}] \label{th:msf}
Consider a good binary linear code ensemble $[\C]$ of rate $R$ whose
transmission takes over $Q$ BISOM parallel channels. Assume that the coded
symbols are randomly and independently assigned to these channels, such that
each bit is transmitted over Channel $j$ with a-priori probability $\tau_j$,
for $j=0,\dots,Q-1$, where $\tau_j>0$ referred to assignment rate and
$\sum_{j=0}^{Q-1}\tau_j=1$. Let $C_j$ and $\gamma_j$ be the channel capacity
and the Bhattacharyya parameter of Channel $j$, and $c_{P}$ and $\xi_{P}$ be
the restriction UB code threshold and the SF distance defined in (\ref{eq:cp})
and (\ref{eq:xip}), respectively. Then, if the average channel capacity and
Bhattacharyya parameter over the $Q$ parallel channels satisfy
\begin{align}
\sum_{j=0}^{Q-1} \tau_j \gamma_j<\exp(-c_{P}) \quad \text{and} \quad
\sum_{j=0}^{Q-1} \tau_j C_j
>R+\xi_{P}, \label{eq:msf-nr-cond}
\end{align}
the average ML decoding word error probability decays to zero as the
codeword length approaches infinity.
\end{theorem}

Different from the condition (\ref{eq:sth-cd}) for the simple code threshold,
the above MSF reliable channel region (\ref{eq:msf-nr-cond}) requires
satisfying two constraints associated with two pairs of channel/code
parameters. Hence, this calculation is complex, in particular, when the
condition (\ref{eq:msf-nr-cond}) is employed with some practical communication
schemes.

\subsection{Some Useful Inequalities}\label{app:eqs}
\begin{proposition}\label{p:ine}
$\qquad 1+b^2\le 2^b\le 1+b$  for $0\le b \le 1$, where the two equalities hold
simultaneously when $b=0$ and $b=1$.

\begin{proof}
Let $g(b)=1+b^2-2^{b}$. We check the derivatives of $g(b)$ in the
region $b\in [0,\,1]$ as follows
\begin{align*}
g'(b)&=2b-\ln 2\cdot 2^b,\\
g''(b)&=2-(\ln 2)^2 \cdot 2^b\ge 2\cdot [1-(\ln 2)^2]>0, \quad
\text{for}~0\le b\le 1.
\end{align*}
This means that $g(b)$ is strictly convex. Note that $g(0)=g(1)=0$.
Thus, $g(b)< 0$ for $b\in (0,\,1)$ and the first inequality of
Proposition~\ref{p:ine} holds.

Next, let $f(b)=1+b-2^b$, we have
\begin{align*}
f'(b)&=1-\ln 2\cdot 2^b, \\
f''(b)&= -(\ln 2)^2\cdot 2^b<0  \qquad \text{for}~0\le b\le 1.
\end{align*}
This implies that $f(b)$ is strictly concave. Since $f(0)=f(1)=0$,
we have $f(b)> 0$ for $b\in (0,\,1)$ and the second inequality of
Proposition~\ref{p:ine} holds.
\end{proof}
\end{proposition}

\begin{proposition}\label{p:ineq}
$\qquad 2^{-a}+2^{-1/a}\le 1$ for $a \ge 0,$ where the equality
holds when $a=0$, $1$, and $\infty$.

\begin{proof}
Consider $f(a)=2^{-a}+2^{-1/a}$. Due to the symmetry property of the
function $f(\cdot)$, i.e., $f(a)=f(1/a)$, we only need to prove
$f(a)\le 1$ for $0\le a\le 1.$

Note that $f(a)$ is differentiable and bounded ($0 \le f(a) < 2$).
Thus, the maximum value of $f(a)$ is either at a boundary point or
at a stationary point. Now we check all such kind of points. For
boundary points, clearly,
\begin{align}
f(a)=1 \quad \text{when}~a=0,1. \label{eq:pL1}
\end{align}
By the definition, the stationary point $s$ satisfies
\begin{align} f'(a)\,\big|_{a=s}=-\ln 2\cdot
(2^{-s}-\frac{1}{s^2}2^{-1/s})=0.
\end{align}
This implies $2^{-1/s}=s^2\cdot2^{-s}$. Thus, for a arbitrary
stationary point $s$, we have
\begin{align}
f(s)&=2^{-s}+2^{-1/s}=2^{-s}+s^2\cdot2^{-s}=2^{-s}(1+s^2)<1~~~\text{for}~0<s<1
\label{eq:pL2}
\end{align}
where the inequality follows from Proposition~\ref{p:ine}. Hence,
(\ref{eq:pL1}) and (\ref{eq:pL2}) imply the desired result.
\end{proof}
\end{proposition}

\subsection{Proof of Lemma~\ref{lem:BI}}\label{app:BI}
The proof of Lemma~\ref{lem:BI} is based on Jensen's inequality
\cite{cover:thomas} and Proposition~\ref{p:ine},~\ref{p:ineq} in
Appendix~\ref{app:eqs}.

\begin{proof}
First, we prove $B\le C$. Note that
\begin{align*}
C-B =&-\E\left\{\log_2\left[\frac{p(Y|0)+p(-Y|0)}{p(Y|0)}\cdot
2^{-\sqrt{p(-Y|0)/p(Y|0)}}\Biggl|\,0\right]\right\}.
\end{align*}
Following Jensen's inequality \cite{cover:thomas}, we have
\begin{align}
C-B &\ge -\log_2\left\{\sum_{y}[p(y|0)+p(-y|0)]
\cdot 2^{-\sqrt{p(-y|0)/p(y|0)}}\right\} \nonumber\\
&=-\log_2\Biggl\{p(y=0|0)+\sum_{y>0}\bigl[p(y|0)+p(-y|0)\bigr] \cdot
\left[2^{-\sqrt{p(-y|0)/p(y|0)}}+2^{-\sqrt{p(y|0)/p(-y|0)}}\right]\Biggr\}.
\label{eq:c-b}
\end{align}
By using Proposition~\ref{p:ineq}, we have
\begin{align*}
2^{-\sqrt{p(-y|0)/p(y|0)}}+2^{-\sqrt{p(y|0)/p(-y|0)}}\le 1
\end{align*}
Hence, (\ref{eq:c-b}) can be bounded as
\begin{align}
C -B
&\ge -\log_2 \Bigl\{p(y=0|0)+\sum_{y>0}[p(y|0)+p(-y|0)]\Bigr\}\nonumber\\
&=-\log_2 1 \nonumber\\
&=0. \label{eq:c-b2}
\end{align}
Now, we prove $B\ge R_0$. Following Proposition~\ref{p:ine}, we have
\begin{align*}
\quad \gamma &\le \log_2 (\gamma+1)\qquad~ \text{for}~0\le \gamma
\le 1,
\end{align*}
By the definition of the cutoff rate in (\ref{eq:r0d}), we can bound the
Bhattacharyya rate as
\[B=1-\gamma \ge 1-\log_2 (\gamma+1)=R_0.\]
By combing (\ref{eq:c-b2}), we have the desired result.
\end{proof}

\subsection{Proof of Lemma~\ref{lem:rate}}\label{app:rate}
The proof of Lemma~\ref{lem:rate} is by contradiction.

\begin{proof}
We assume that (\ref{eq:lm-rate}) does not hold. Then, there exists
a positive $\epsilon_0$ such that
\begin{align}
R>1-\exp(-\ubth)+\epsilon_0. \label{eq:lmcon}
\end{align}
Let's consider a binary erasure channel (BEC) with erasure
probability $p=\exp(-\ubth)-\epsilon_0$. Then, the channel capacity
and the Bhattacharyya parameter are
\begin{align*}
C(p)& \triangleq 1-p =1-\exp(-\ubth)+\epsilon_0 \quad \text{and}
\quad \gamma(p)\triangleq p =\exp(-\ubth)-\epsilon_0.
\end{align*}
Since $\epsilon_0>0$, the UB reliable communication condition
(\ref{eq:condc0}) is satisfied, i.e., $\gamma(p)<\exp(-\ubth)$.
Hence, the decoding error probability approaches $0$ as
$N\rightarrow 0$. Now, the converse to Shannon's channel coding
theorem \cite{cover:thomas} implies
\begin{align*}
R\le C(p)=1-\exp(-\ubth)+\epsilon_0,
\end{align*}
which contradicts (\ref{eq:lmcon}).
\end{proof}

\subsection{Proof of Theorem~\ref{th:cth}}\label{app:cth}

For the proof of Theorem~\ref{th:cth} we proceed in the following two steps. We
first show the existence of $\cth$ by using Lemma~\ref{lem:rate}. Next, we
prove the main result based on the MSF reliable channel region theorem (see
Appendix~\ref{app:msf}) and Lemma~\ref{lem:BI}.

\begin{proof}
Let
\[\Omega\triangleq \{P:~1-\exp(-c_{P}) \ge R+\xi_{P}~\text{and}~0\le P <0.5\}.\]
To prove the existence of $\cth$, we need to show that the set
$\Omega$ is not empty. Let's consider a particular weight partition
that $P=0$. In this case, Equations (\ref{eq:cp}) and (\ref{eq:xip})
imply $c_{P}=\ubth$ and $\xi_{P}=0$. Moreover, Lemma~\ref{lem:rate}
leads to
\[
1-\exp(-c_{P}) \ge R+ \xi_{P} \quad \text{for} \quad P=0.
\]
Hence, the set $\Omega\neq\emptyset$ and $\cth$ is well defined.

Next, we prove that, for $\forall~P\in\Omega$,
$P_{W}(\bar{\gamma},N)\stackrel{N}{\longrightarrow}0$ if
\begin{align}
\bar{\gamma}<\exp(-c_{P})\quad. \label{eq:thc}
\end{align}
The condition (\ref{eq:thc}) implies
\begin{align*}
1-\bar{\gamma}\ge 1-\exp(-c_{P}) \ge R+ \xi_{P}^{[\C]} \quad
\text{for}~P\in\Omega
\end{align*}
where the second inquilinity follows from the definition of $\Omega$.  On the
other hand, Lemma~\ref{lem:BI} implies that the average channel capacity
\begin{align*}
\bar{C}\triangleq\sum_{j=0}^{Q-1} \tau_j C_j \ge \sum_{j=1}^{Q-1}
(1-\tau_j \gamma_j)=1-\bar{\gamma}.
\end{align*}
Therefore, $$\bar{C}>R+ \xi_{P}.$$ By combining (\ref{eq:thc}) and
Theorem~\ref{th:msf}, we have the desired result
$P_{W}(\bar{\gamma},N)\stackrel{N}{\longrightarrow}0$.
\end{proof}

\subsection{Discussion on Reducing Decoding Delay} \label{app:rtau}

Here, we propose ``early stopping''  and ``threshold adjusting'', which in
practice neutralize the effect of decoding delay.

Without of loss generality, we assume that Node $1$ is a reliable node and
scheduled to send the message in slot~$1$. The key of early stopping is that
Node $1$ does not need to listen to the message during the whole slot
$[0,\,\tau_0T]$, instead, Node $1$ can stop listening early and begin to decode
after it receives enough information. Let $\theta_{0,1}$ be the
sender-to-helper instantaneous SNR of Node $1$. By the reliable node definition
(see Sec.~\ref{subsec:bs}), we have $\theta_{0,1}>\chi(\tau_0)$. Hence there
exists a effective listening period $\tau'$ satisfying
\begin{align}
\theta_{0,1}>\chi(\tau')>\chi(\tau_0). \label{eq:eft}
\end{align}
Equations (\ref{eq:eft}) and (\ref{eq:punctured}) imply that $\tau'$ satisfies
\[\frac{1-\exp(-\cth)}{1-\exp(-\theta_{0,1})}<\tau'<\tau_0.\]
Node $1$ begins to decode after receiving $\tau'N$ bits.
Theorem~\ref{th:pcth} implies that the ``early stopping'' rule can
guarantee ML decoding successful at Node $1$. Now, the required
extra time is $\tau_D+\tau'-\tau_0$. Clearly, the proposed strategy
helps in reducing decoding delay latency.

When $\tau_D\ll \tau_0$, an alternative method is threshold adjusting. Here, we
reset the punctured code threshold to be $\chi(\tau_0-\tau_D)$. We say that
Node $1$ is a reliable node (scheduled to send the message in slot $1$) when
the sender-to-helper instantaneous SNR
\begin{align}
\theta_{0,1}>\chi(\tau_0-\tau_D).
\end{align}
Node $1$ listens to the message during the slot $[0,\,(\tau_0-\tau_D)T]$ and
begins to decode after receiving $(\tau_0-\tau_D)N$ bits. Now, there is still
$\tau_DT$ in slot $0$ for decoding process.

\subsection{Proof of Equation (\ref{eq:cofer2})} \label{app:cof}
Here, we provide the intermediate steps for deriving
(\ref{eq:cofer2}).

\begin{proof}
Let $$x=\exp(-\nu_{0,2})~\text{and}~y=\exp(-\nu_{1,2}).$$ Recall that fading
power gains $\nu_{1,2}$ and $\nu_{0,2}$ are independent exponentially
distributed with mean $1$. Hence random variables $x$ and $y$ are independent
uniformly distributed over $(0,1]$. By the definition (\ref{eq:funcG}), we have
\begin{align}
 {\mathcal G}(2,\{1\},\bm{\snr})
&={\rm P}\bigl\{\tau_0\gamma_0+\tau_1\gamma_1\ge \exp(-\cth)\bigr\}
\nonumber\\
&=1-{\rm P}\bigl\{\tau_0\gamma_0+\tau_1\gamma_1< \exp(-\cth)\bigr\}
\nonumber\\
&=1- {\rm P}\bigl\{\tau_0\gamma_0+\tau_1< \exp(-\cth)\bigr\}-{\rm
P}\bigl\{\tau_0\gamma_0+\tau_1\ge
\exp(-\cth),\,\tau_0\gamma_0+\tau_1\gamma_1<
\exp(-\cth)\bigr\}\nonumber\\
&=1-{\rm P}\bigl\{\gamma_0< \exp[-\chi(\tau_0)]\bigr\}- {\rm
P}\Bigl\{\gamma_0\ge \exp[-\chi(\tau_0)],\,\gamma_1<
\frac{\exp(-\cth)-\tau_0\gamma_0}{\tau_1}\Bigr\} \label{eq:pcof}
\end{align}
where the third equality is due to $0\le\gamma_1\le1$, and the last
equality is based on the definition of punctured code threshold
(\ref{eq:punctured}). Note that
\begin{align*}
\gamma_0=\exp(-\nu_{0,2}\snr_{0,2})=x^{\snr_{0,2}}\\
\gamma_1=\exp(-\nu_{1,2}\snr_{1,2})=y^{\snr_{1,2}}.
\end{align*}
Hence, the second term of (\ref{eq:pcof}) is given by
\begin{align}
 {\rm P}\bigl\{\gamma_0< \exp[-\chi(\tau_0)]\bigr\}
&={\rm P}\bigl\{x<\omega \bigr\}=\omega \label{eq:pcof1}
\end{align}
where $\omega=\exp\bigl[-\chi(\tau_0)\cdot\snr_{0,2}^{-1}\bigr].$
Next, the third term of (\ref{eq:pcof}) can be rewritten as
\begin{align}
{\rm P}\Bigl\{\gamma_0\ge \exp[-\chi(\tau_0)],\,\gamma_1<
\frac{\exp(-\cth)-\tau_0\gamma_0}{\tau_1}\Bigr\} &={\rm P}\Biggl\{x\ge
\omega,\,y <
\Biggl[\frac{\exp(-\cth)-\tau_0x^{\snr_{0,2}}}{\tau_1}\Biggr]^{1/\snr_{1,2}}
\Biggr\}\nonumber\\
&
=\int_{\omega}^{1}\Biggl[\frac{\exp(-\cth)-\tau_0x^{\snr_{0,2}}}{\tau_1}\Biggr]^{1/\snr_{1,2}}\,dx
\label{eq:pcof2}
\end{align}
Combining (\ref{eq:pcof}), (\ref{eq:pcof1}), and (\ref{eq:pcof2})
together, we have the desired result (\ref{eq:cofer2}).
\end{proof}

\subsection{Proof of Theorem~\ref{th:lim}} \label{app:bound}

The intuition of the proof of Theorem~\ref{th:lim} is described as follows.
Since $Q$ is a positive integer, we derive the result based on an induction
method. First, we consider the $Q=2$ case and derive Lemma~\ref{lem:ab2} based
on a more general setting. Next, we prove the main result of
Theorem~\ref{th:lim}. We show that if the hypothesis (\ref{eq:lm2-p0}) holds
for the integer $Q=j-1$, it is true for the next greater value $Q=j$.

\begin{lemma}\label{lem:ab2}
Consider two independent random variables $\phi_1$ and $\phi_2$, where $0 \le
\phi_m\le 1$ and the probability distribution of $\phi_m$ is a function of
$\lambda_m$, for $m=1,2$. Assume that
\begin{align}
\lim_{\lambda_1 \rightarrow\infty} \lambda_1\,{\rm
P}[\phi_1>c]=f(c),~~~
\limsup_{\lambda_2\rightarrow\infty}\;\lambda_2\,{\rm
P}[\phi_2>c]\le h(c),
\end{align}
where $f(c)$ and $h(c)$ are monotone decreasing and integrable, and
$f'(c)$ is integrable. If
$$0<\tau,\,1-\tau<c,$$
then
\begin{align}
\limsup_{\lambda_1,\lambda_2\rightarrow\infty}\;&\lambda_1\lambda_2\,
{\rm P}[\tau \phi_1+(1-\tau) \phi_2>c] \le-\int_{q}^1
h\Bigl(\frac{c-\tau z}{1-\tau}\Bigr)f'(z)dz \label{eq:lem3}
\end{align}
where $q=[c-(1-\tau)]/\tau.$

\begin{proof}[Lemma~\ref{lem:ab2}]
The proof follows the approach in \cite{lw-IT}. First, let ${\mathcal Z}=\{z_0,
\dots ,z_L\}$ for some finite $L$ be any partition of the interval $[q,\,1]$
with $z_0 =q$ and $z_L =1$. Next we obtain an outer bound on the event $\{\tau
\phi_1+(1-\tau) \phi_2>c\}$ as
\begin{align}
\{\tau \phi_1+(1-\tau) \phi_2>c\} & \subseteq \bigcup^{L}_{i=1}
\Biggl\{\{z_{i-1}\le \phi_1 < z_i\}\cap \Bigl\{\phi_2> \frac{c-\tau
z_{i}}{1-\tau}\Bigr\}\Biggr\}
\end{align}
Since $\phi_1$ and $\phi_2$ are independent, we have
\begin{align}
{\rm P}\Biggl[z_{i-1}\le \phi_1 < z_{i},~\phi_2>\frac{c-\tau
z_{i}}{1-\tau}\Biggr] =&\,\Bigl\{{\rm P}[\phi_1>z_{i-1}]-{\rm
P}[\phi_1>z_{i}]\Bigr\}\,{\rm P}\left[\phi_2>\frac{c-\tau z_{i}}{1-\tau}\right]
\end{align}
Hence,
\begin{align}
\limsup_{\lambda_1,\lambda_2\rightarrow\infty}\lambda_1\lambda_2\,{\rm P}[\tau
\phi_1+(1-\tau) \phi_2>c] &\le
\sum^{L}_{i=1}\limsup_{\lambda_1,\lambda_2\rightarrow\infty}~\lambda_1\lambda_2\,{\rm
P}\Biggl[z_{i-1}\le \phi_1 <
z_{i},~\phi_2>\frac{c-\tau z_{i}}{1-\tau}\Biggr] \nonumber \\
&\le \sum^{L}_{i=1}\limsup_{\lambda_1,\lambda_2\rightarrow\infty}
\lambda_1\Bigl\{{\rm P}[\phi_1>z_{i-1}]-{\rm
P}[\phi_1>z_{i}]\Bigr\}\cdot\lambda_2{\rm P}\Biggl[\phi_2>\frac{c-\tau
z_{i}}{1-\tau}\Biggr] \nonumber \\
&=\sum^{L}_{i=1}\bigl[f(z_{i-1})-f(z_{i})\bigr]h\Bigl(\frac{c-\tau
z_{i}}{1-\tau}\Bigr). \label{eq:lem1}
\end{align}
Note that (\ref{eq:lem1}) holds for all partitions ${\mathcal Z}$ of the
interval $[q,\,1]$;  and $f(c),$ $f'(c),$ and $h(c)$ are all integrable, the
supremum of the right-hand side of (\ref{eq:lem1}) becomes the integral in
(\ref{eq:lem3}).
\end{proof}
\end{lemma}

Now, we prove Theorem~\ref{th:lim} following the induction method.

\begin{proof}[Theorem~\ref{th:lim}]
We first check the $Q=1$ case. Note that $\tau_1=1$ in this case, thus,
(\ref{eq:u1}) implies the hypothesis (\ref{eq:lm2-p0}) holds for $Q=1$.

Next we assume that the hypothesis (\ref{eq:lm2-p0}) holds for $Q=j-1$ and
consider $Q=j$. Let
\begin{align*}
 \tau_m'
&=\frac{\tau_m}{1-\tau_j} \quad \text{for}~m=1,\dots,j-1,
\end{align*}
and
\begin{align*}
f(c)&=-\ln c\\
h(c)&=\frac{1}{(j-1)!}\prod_{m=1}^{j-1}\ln\frac{\tau_m'}{c-(1-\tau_m')}.
\end{align*}
Since $\tau_m'<1-c$, $\sum_{m=1}^{j-1}\tau_m'=1$, the induction hypothesis for
$Q=j-1$ implies
\begin{align}
\limsup_{\lambda_1,\dots,\lambda_{j-1}\rightarrow\infty}\prod_{m=1}^{j-1}\lambda_m
\cdot {\rm P}\left[\sum_{m=1}^{j-1}\tau_m' \phi_m >c\right] \le
h(c).
\end{align}
Note that $f(c)$ and $h(c)$ are monotone decreasing and integrable, and $f'(c)$ is integrable.
Then, by Lemma~\ref{lem:ab2},
\begin{align}
\limsup_{\lambda_1,\dots,\lambda_{j}\rightarrow\infty}\;
\prod_{m=1}^{j}\lambda_{m} \, {\rm P}\left[\sum_{m=1}^{j}\tau_m
\phi_m>c\right] = &\
\limsup_{\lambda_1,\dots,\lambda_{j}\rightarrow\infty}\;\prod_{m=1}^{j}\lambda_{m}
\,{\rm P}\left[\tau_j \phi_j+(1-\tau_j)\sum_{m=1}^{j-1}\tau_m'
\phi_m>c\right]
\nonumber \\
\le &\ -\int_{q}^1 h\Bigl(\frac{c-\tau z}{1-\tau}\Bigr)f'(z)dz \nonumber \\
=&\ \frac{1}{(j-1)!}\int_{q}^1 \frac{1}{z}\,\prod_{m=1}^{j-1} \ln
\frac{\tau_m}{c-z\tau_j-(1-\tau_j-\tau_m)}dz \label{eq:lm2-p1}
\end{align}
where $$q=\frac{c-(1-\tau_j)}{\tau_j}.$$ Since $-\ln z$ is a convex
function , Jensen's inequality implies that
\begin{align}
\ln \frac{\tau_m}{c-z\tau_j-(1-\tau_j-\tau_m)}
&=-\ln\Bigr[\frac{z\tau_j-c+(1-\tau_j)}{1-c}\cdot\frac{\tau_m-(1-c)}{\tau_m}+\frac{\tau_j-z\tau_j}{1-c}\cdot 1\Bigl]\nonumber\\
&\le -\frac{z\tau_j-c+(1-\tau_j)}{1-c} \ln \frac{\tau_m-(1-c)}{\tau_m} -\frac{\tau_j-z\tau_j}{1-c} \ln 1\nonumber\\
&= \frac{z-q}{1-q}\ln \frac{\tau_m}{c-(1-\tau_m)} \qquad
\text{for}~~q\le z \le 1,~1\le m \le j-1.
\end{align}
Hence,
\begin{align}
\int_{q}^1 \frac{1}{z}\,\prod_{m=1}^{j-1} \ln \frac{\tau_m}{c-z\tau_j-(1-\tau_j-\tau_m)}dz \le
\left[\prod_{m=1}^{j-1}\ln \frac{\tau_m}{c-(1-\tau_m)}\right] \int_{q}^1 \frac{1}{z}\
\Biggl[\frac{z-q}{1-q}\Biggr]^{j-1} dz  . \label{eq:lm2-p2}
\end{align}
Note that $1/z$ and $[(z-q)/(1-q)]^{j-1}$ are, respectively,
monotonically decreasing and increasing in $z$. Chebyshev integral
inequality \cite{mitr} implies
\begin{align}
\int_{q}^1 \frac{1}{z}\ \Biggl[\frac{z-q}{1-q}\Biggr]^{j-1} dz &\le
\frac{1}{1-q}\int_{q}^1
\Biggl[\frac{z-q}{1-q}\Biggr]^{j-1} dz \cdot \int_{q}^1 \frac{1}{z}dz \nonumber\\
&=-\frac{\ln q}{j}  . \label{eq:lm2-p3}
\end{align}
Finally, by combining (\ref{eq:lm2-p1}), (\ref{eq:lm2-p2}), and (\ref{eq:lm2-p3}), we have the
desired result
\begin{align}
\limsup_{\lambda_1,\dots,\lambda_{j}\rightarrow\infty}\;
\prod_{m=1}^{j}\lambda_{m} \, {\rm P}\left[\sum_{m=1}^{j}\tau_m
\phi_m >c\right] &\le \ \frac{1}{j!}\left[\prod_{m=1}^{j}\ln
\frac{\tau_m}{c-(1-\tau_m)}\right].
\end{align}
Since both the baseline case ($Q=1$) and the inductive step ($Q=j$) satisfy
(\ref{eq:lm2-p0}), we conclude that the hypothesis (\ref{eq:lm2-p0}) holds for
all $Q$.
\end{proof}

\section*{ACKNOWLEDGMENT}

The authors would like to thank anonymous reviewers for helpful
suggestions to improve the presentation of the paper.

\renewcommand{\baselinestretch}{1.6}
\bibliographystyle{IEEEtran}
\bibliography{coop}

\end{document}